\documentclass[prd,aps,twocolumn,nofootinbib,10pt]{revtex4-1}
\usepackage{amssymb,amsmath,amsfonts,amsbsy,epsfig,graphicx}
\usepackage[dvipsnames]{xcolor}
\usepackage{braket}
\definecolor{lgray}{gray}{0.35}
\usepackage[colorlinks=true,urlcolor=Gray,linkcolor=lgray,
citecolor=Gray,
             pdfpagelabels=true,hypertexnames=true,
            plainpages=false,naturalnames=false,
             ]{hyperref}

\newcommand{\be}{\begin{equation}}
\newcommand{\ee}{\end{equation}}
\newcommand{\bea}{\begin{eqnarray}}
\newcommand{\eea}{\end{eqnarray}}
\newcommand{\nn}{\nonumber}

\def\X{{\mathcal X}}
\def\Y{{\mathcal Y}}

\def\p{{\bf p}}
\def\q{{\bf q}}
\def\k{{\bf k}}
\def\r{{\bf r}}
\def\x{{\bf x}}
\def\y{{\bf y}}
\def\z{{\bf z}}

\begin{document}

\title{Landscape tomography through primordial non-Gaussianity}

\date{October 17, 2018}

\author{
Xingang Chen$^{a}$, Gonzalo A. Palma$^{b}$, Walter Riquelme$^{b}$, Bruno Scheihing H.$^{b}$ and Spyros Sypsas$^{b}$
}

\affiliation{
$^{a}$Institute for Theory and Computation, Harvard-Smithsonian Center for Astrophysics, 60 Garden Street, Cambridge, MA 02138, USA. \\
$^{b}$Grupo de Cosmolog\'ia y Astrof\'isica Te\'orica, Departamento de F\'{i}sica, FCFM, \mbox{Universidad de Chile}, Blanco Encalada 2008, Santiago, Chile.}

\begin{abstract}

In this paper, we show how the structure of the landscape potential of the primordial Universe may be probed through the properties of the primordial density perturbations responsible for the origin of the cosmic microwave background anisotropies and the large-scale structure of our Universe. Isocurvature fields ---fields orthogonal to the inflationary trajectory--- may have fluctuated across the barriers separating local minima of the landscape potential during inflation. We analyze how this process could have impacted the evolution of the primordial curvature perturbations. If the typical distance separating consecutive minima of the landscape potential and the height of the potential barriers are smaller than the Hubble expansion rate parametrizing inflation, the probability distribution function of isocurvature fields becomes non-Gaussian due to the appearance of bumps and dips associated with the structure of the potential. We show that this non-Gaussianity can be transferred to the statistics of primordial curvature perturbations if the isocurvature fields are coupled to the curvature perturbations. The type of non-Gaussian structure that emerges in the distribution of curvature perturbations cannot be fully probed with the standard methods of polyspectra; instead, the probability distribution function is needed. The latter is obtained by summing all the $n$-point correlation functions. To substantiate our claims, we offer a concrete model consisting of an axionlike isocurvature perturbation with a sinusoidal potential and a linear derivative coupling between the isocurvature and curvature field. In this model, the probability distribution function of the curvature perturbations consists of a Gaussian function with small superimposed oscillations reflecting the isocurvature axion potential.

\end{abstract}

\maketitle

\section{Introduction}

The search for primordial non-Gaussianity (NG) has been guided by our ability to make predictions within the inflationary paradigm~\cite{Guth:1980zm, Linde:1981mu, Albrecht:1982wi, Starobinsky:1980te, Mukhanov:1981xt}. The simplest models of inflation predict that the main departures from Gaussianity are to be found in the form of small, but nonvanishing, three-point correlation functions of the primordial comoving curvature perturbation $\zeta$~\cite{Gangui:1993tt, Komatsu:2001rj, Acquaviva:2002ud, Maldacena:2002vr}. However, interactions involving the inflaton and other fields could enhance the amplitude of the three-point or higher point correlation functions (see \cite{Bartolo:2004if,Liguori:2010hx,Chen:2010xka,Wang:2013eqj} for reviews). These interactions could be self-interactions of the inflaton or interactions of the inflaton with other degrees of freedom. While current cosmic microwave background (CMB) constraints on the bispectrum are consistent with Gaussian statistics~\cite{Ade:2015ava}, it is possible that the method of three-point or higher-point correlation functions do not constitute the most efficient parametrization of primordial NG hidden in the data~\cite{Komatsu:2003fd, Buchert:2017uup}.

Multifield \cite{Enqvist:2004bk, Lyth:2005fi, Seery:2005gb, Rigopoulos:2005ae, Alabidi:2005qi, Battefeld:2007en, Byrnes:2008wi, Byrnes:2009qy, Battefeld:2009ym, Elliston:2011et, Mulryne:2011ni, McAllister:2012am, Byrnes:2012sc, Bjorkmo:2017nzd, Langlois:1999dw, Gordon:2000hv, GrootNibbelink:2000vx, GrootNibbelink:2001qt, Amendola:2001ni, Bartolo:2001rt, Wands:2002bn, Tsujikawa:2002qx, Byrnes:2006fr, Choi:2007su, Cremonini:2010sv, Cremonini:2010ua, Achucarro:2016fby} and quasi-single-field \cite{Chen:2009we,Chen:2009zp, Baumann:2011nk, Chen:2012ge, Sefusatti:2012ye, Norena:2012yi, Noumi:2012vr, Gong:2013sma, Emami:2013lma, Kehagias:2015jha, Arkani-Hamed:2015bza,  Dimastrogiovanni:2015pla,Chen:2015lza, Chen:2016cbe,Lee:2016vti,Chen:2016qce, Meerburg:2016zdz,Chen:2016uwp,Chen:2016hrz, An:2017hlx,Iyer:2017qzw,An:2017rwo, Kumar:2017ecc, Franciolini:2017ktv,Tong:2018tqf, MoradinezhadDizgah:2018ssw, Saito:2018xge,Franciolini:2018eno,Chen:2018sce} models of inflation constitute a particularly useful framework to study the generation of large primordial NG. The interaction between curvature perturbations and isocurvature fields are known to offer NG departures by enhancing the amplitude of the three-point correlation function of $\zeta$. For example, a light \cite{Enqvist:2004bk, Lyth:2005fi, Seery:2005gb, Rigopoulos:2005ae, Alabidi:2005qi, Battefeld:2007en, Byrnes:2008wi, Byrnes:2009qy, Battefeld:2009ym, Elliston:2011et, Mulryne:2011ni, McAllister:2012am, Byrnes:2012sc, Bjorkmo:2017nzd, Langlois:1999dw, Gordon:2000hv, GrootNibbelink:2000vx, GrootNibbelink:2001qt, Amendola:2001ni, Bartolo:2001rt, Wands:2002bn, Tsujikawa:2002qx, Byrnes:2006fr, Choi:2007su, Cremonini:2010sv, Cremonini:2010ua, Achucarro:2016fby}
or massive \cite{Chen:2009we,Chen:2009zp, Baumann:2011nk, Chen:2012ge, Achucarro:2012yr, Sefusatti:2012ye, Norena:2012yi, Noumi:2012vr, Gong:2013sma, Emami:2013lma, Kehagias:2015jha, Arkani-Hamed:2015bza,  Dimastrogiovanni:2015pla,Chen:2015lza, Chen:2016cbe,Lee:2016vti,Chen:2016qce, Meerburg:2016zdz,Chen:2016uwp,Chen:2016hrz, An:2017hlx,Iyer:2017qzw,An:2017rwo, Kumar:2017ecc, Franciolini:2017ktv,Tong:2018tqf, MoradinezhadDizgah:2018ssw, Saito:2018xge,Franciolini:2018eno,Chen:2018sce, Tolley:2009fg, Achucarro:2010jv, Achucarro:2010da, Cespedes:2012hu} isocurvature field $\psi$ may affect the dynamics of curvature perturbations during horizon crossing and/or at superhorizon scales due to a special type of derivative coupling of the form
\be
\mathcal L_{\rm int} \propto \dot \zeta \psi . \label{der-coupl}
\ee
This interaction appears in the Lagrangian describing the dynamics of fluctuations when the inflationary path experiences a turn in the multiple-field target space \cite{Gordon:2000hv, GrootNibbelink:2000vx, GrootNibbelink:2001qt}. This derivative coupling, or other types of couplings appearing at higher order, may communicate different kinds of nonlinearities present in the multiple-field space to the curvature perturbation $\zeta$.

\begin{figure}[t!]
\includegraphics[scale=0.35]{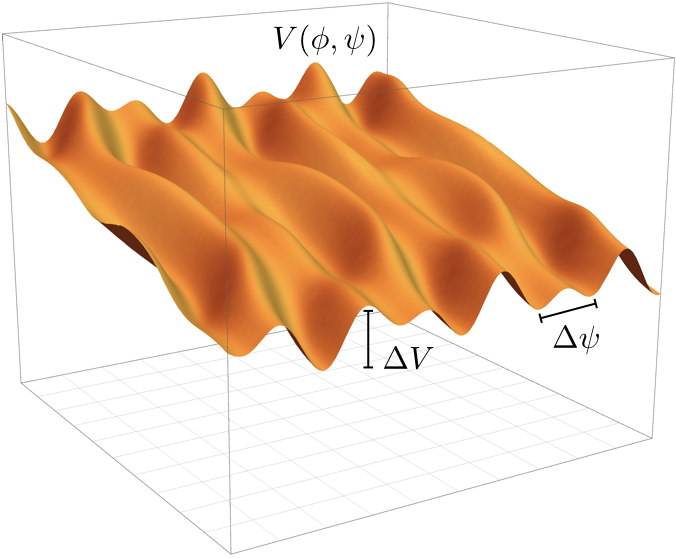}
\caption{Illustration of a situation where the isocurvature mode would be able to fluctuate with an amplitude that would traverse several minima of the landscape. We are interested in situations where both the characteristic distance between close minima $\Delta \psi$ and the characteristic height of the potential barrier $\Delta V^{1/4}$ are smaller than $H$, which is the typical amplitude of fluctuations around horizon crossing.}
\label{fig:FIG_landscape}
\end{figure}

In this work we extend this mechanism [involving the derivative coupling of Eq.~(\ref{der-coupl})] and show that the probability distribution function (PDF) of primordial curvature perturbations may inherit a novel class of non-Gaussianity. It relies on the existence of an isocurvature field $\psi$ that acquires non-Gaussian statistics through its own self-interactions~\cite{Palma:2017lww}, which are transferred to curvature perturbations on superhorizon scales. The self-interactions of the isocurvature field are determined by a given potential $\Delta V(\psi)$ characterized for having a rich structure within a field range smaller than $H$, the Hubble expansion rate of the Universe during inflation. More precisely, the potential for the isocurvature field is assumed to have several local minima and barriers separated by a characteristic distance $\Delta \psi$. We also assume that the scale of the barrier height $\Delta V^{1/4}$  is smaller than $H$ (see Fig.~\ref{fig:FIG_landscape}). The resulting picture is reminiscent of the string landscape~\cite{Susskind:2003kw} (see Ref.~\cite{Linde:2015edk} for a recent discussion on certain implications of the landscape picture for cosmology).
In this situation the isocurvature field will be able to vigorously jump and diffuse across the \mbox{barriers}. As a result, its PDF will be such that, after horizon crossing, it becomes more probable to measure a given amplitude of $\psi$ that coincides with a local minimum of the landscape potential. As a consequence, the PDF of $\zeta$ will inherit a similar non-Gaussian profile.

The type of non-Gaussianity transferred to the curvature fluctuations cannot be fully parametrized with low $n$-point correlation functions such as three- and four-point spectra. Instead, to describe the type of NG that we encounter here, one needs to take into account a larger set of $n$-point correlation functions, revealing information about the nontrivial structure of the  $\zeta$ PDF. This means that, in order to constrain or unveil the effects discussed in this article, one needs methods different from those often employed to test NG. More to the point, to test this class of NG one needs to constrain the entire shape of the PDF for $\zeta$. Methods to constrain the primordial PDF have already been introduced in the past~\cite{Komatsu:2003fd, Buchert:2017uup} but understandably have received much less attention than those useful to constrain the three- and four-point functions. In this work, we will not focus on such methods. Instead, we will study the mechanism underlying the emergence of this type of non-Gaussianity in a specific well-motivated example, involving an axion field.

The main result that we derive here is the PDF for $\zeta$ in the particular case that $\Delta V(\psi) = \Lambda^4 \left[ 1 - \cos (\psi / f) \right]$. This PDF is given by Eq.~(\ref{PDF-full}) and plotted in Figs.~\ref{fig:FIG-PDF}-\ref{fig:FIG-PDF-2} for various choices of parameters. In a companion Letter~\cite{Chen:2018brw}, we argue that this result can be extended to more general potentials $\Delta V(\psi)$ and show that the shape of the PDF is entirely determined by the potential $\Delta V(\psi)$ for $\psi$. Together with~\cite{Chen:2018brw}, the results of this article imply that, in principle, this type of NG is able to provide information about the shape of a section of the landscape at the time of horizon crossing: a landscape tomography. Whereas, by restricting oneself to the moments of this PDF (i.e., the $n$-point correlation functions) one completely misses this structure.

If observations ever confirm the existence of this type of non-Gaussianity, we would have a concrete way of identifying a new energy scale parametrizing the field range of the landscape potential involving degrees of freedom additional to the inflaton. Indeed, the shapes in momentum space of the $n$-point functions leading to the PDF are of the local type. These shapes cannot be obtained through self-interactions arising in single-field inflation since they would violate Maldacena's consistency relation~\cite{Maldacena:2002vr} (or more generally, the soft limit theorems in single-field inflation~\cite{Creminelli:2004yq, Cheung:2007sv, Tanaka:2011aj, Creminelli:2012ed, Assassi:2012zq, Pajer:2013ana, Joyce:2014aqa, Bravo:2017wyw, Finelli:2017fml, Bravo:2017gct}). Therefore, tomographic non-Gaussianity can only be attributed to degrees of freedom (fluctuating across the landscape) whose effect on the spectra cannot be integrated out.

The present article has been organized as follows: We begin in Sec.~\ref{sec:mechanism} by describing the general mechanism by which non-Gaussianity may arise in the distribution of curvature perturbation due to isocurvature self-interactions. In Sec.~\ref{sec:stage} we introduce the specific perturbation theory (and describe the in-in formalism) that will be used to compute $n$-point correlation functions. In Sec.~\ref{sec:linearth} we discuss the linear theory that arises in the absence of the potential $\Delta V(\psi)$, but in the presence of the derivative interaction, coupling together the curvature and isocurvature perturbations, responsible for the transfer of the NG statistics. Then, in Sec.~\ref{sec:npoints} we proceed to compute all $n$-point correlation functions due to the self-interactions of the isocurvature field in a nonperturbative manner. In Sec.~\ref{sec:tom-NG} we use the general expression obtained in the previous section to derive the probability distribution function incorporating the non-Gaussian structure reflecting the axionlike potential. We discuss our results and provide our concluding remarks in Sec.~\ref{sec:conclusions}. In order to alleviate the exposition, we have left some (important) material for the appendices: In Appendix~\ref{sec:concrete-example}, we offer a concrete example of a specific UV complete model from where the action for perturbations used in Section~\ref{sec:stage} descents. Last but not least, in Appendices~\ref{sec:details} and~\ref{app_about_I_n} we provide details of some steps omitted in the computations of Secs.~\ref{sec:npoints} and~\ref{sec:tom-NG}, respectively.

\section{Description of the mechanism} \label{sec:mechanism}

It is well known that, in two-field models of inflation with turning trajectories, to quadratic order in the fluctuations, the evolution of the comoving curvature perturbation $\zeta$ coupled to a single isocurvature field $\psi$ is described by the following Lagrangian~\cite{Gordon:2000hv, GrootNibbelink:2000vx, GrootNibbelink:2001qt}:
\bea
\mathcal L (\zeta , \psi) &=&  a^3 \Big[\epsilon  ( \dot \zeta - \alpha \psi ) ^2 - \frac{\epsilon}{a^2}   (\nabla \zeta)^2  \nn \\
&& +  \frac{1}{2} \dot \psi^2 - \frac{1}{2 a^2} (\nabla \psi)^2  -  \frac{1}{2} \mu^2 \psi^2 \Big],\label{Lagrantian-quadratic}
\eea
where $\epsilon \equiv - \dot H / H^2$ is the usual first slow-roll parameter. Here, $\mu$ corresponds to the so-called entropy mass of $\psi$. In the long wavelength limit, $\psi$ satisfies the following equation of motion (obtained after integrating the equation of motion for $\zeta$ once)
\be
\ddot \psi + 3 H \dot \psi + \mu^2 \psi = 0 , \label{superhorizon-eom-psi}
\ee
from where it is possible to read that $\mu$ determines the mass of $\psi$ on superhorizon scales.
Notice that $\zeta$ and $\psi$ interact through the coupling $\alpha$ appearing in the special combination
\be
D_t \zeta \equiv \dot \zeta - \alpha \psi,
\ee
determining the kinetic term of $\zeta$. In general the coupling $\alpha$ depends on time, and its appearance may be understood as the consequence of bends of the inflationary trajectory in the multifield target space (or more precisely, nongeodesic motions in target space)~\cite{GrootNibbelink:2000vx, GrootNibbelink:2001qt,Chen:2009we,Chen:2009zp, Achucarro:2010jv, Achucarro:2010da}.

If the entropy mass vanishes ($\mu = 0$), the field $\psi$ becomes ``ultralight", and the system gains a symmetry given by
\bea
&&\psi \to \psi' = \psi + C , \label{symm-1} \\ \label{symm-2}
&&\zeta \to \zeta' = \zeta + C \int^t \!\!  dt \, \alpha ,
\eea
where $C$ is an arbitrary constant. The consequences of this symmetry were investigated in Ref.~\cite{Achucarro:2016fby}. We summarize the findings of~\cite{Achucarro:2016fby} as follows: First, the symmetry of the Lagrangian (\ref{Lagrantian-quadratic}) under the transformation (\ref{symm-1}) ensures the existence of a constant solution for $\psi$. This can be seen directly in Eq.~(\ref{superhorizon-eom-psi}). This solution, say $\psi_*$, spontaneously breaks the symmetry, and dominates on superhorizon scales. Second, the symmetry of the Lagrangian under the transformation (\ref{symm-2}) implies that the constant solution $\psi_*$ will source the evolution of $\zeta$ on superhorizon scales. Concretely, if for simplicity we assume that $\alpha$ is nearly constant, on superhorizon scales one finds:
\be
\zeta \simeq \frac{\alpha}{H} \psi_* \Delta N . \label{zeta-psi-N-1}
\ee
If we conveniently identify $\psi_*$ as the value of the field at horizon crossing, then $\Delta N$ corresponds to the number of $e$-folds after that event. A given $n$-point correlation function is then given by:
\be
\langle \zeta^n \rangle \simeq \left( \frac{\alpha}{H} \Delta N \right)^n \langle \psi_*^n \rangle . \label{zeta-n-psi-n}
\ee
In the particular case of $n=2$, we obtain a relation between the power spectrum of $\zeta$ and the power spectrum of $\psi$:
\be
P_{\zeta} \simeq \frac{\alpha^2 \Delta N^2}{H^2} P_{\psi} . \label{Pzeta-Ppsi}
\ee
Moreover, it is possible to show that $\zeta$ has a negligible influence on the evolution of $\psi$, which behaves as a massless field before and after horizon crossing [recall our comments about $\mu$ after Eq.~(\ref{Lagrantian-quadratic})]. This implies that the power spectrum for $\psi$ is given by
\be
P_{\psi} = \frac{H_*^2}{4 \pi^2} ,
\ee
where $H_*$ is the Hubble parameter evaluated at horizon crossing. Now the key issue to stress about Eqs.~(\ref{zeta-n-psi-n}) and~(\ref{Pzeta-Ppsi}) is that the statistics of the field $\zeta$ is completely determined by the statistics of $\psi$. In this case, given that we are only considering a quadratic Lagrangian without higher order self-interactions for $\psi$, the statistics of $\psi$ is found to be Gaussian. Then, the statistics inherited by $\zeta$ is also found to be Gaussian, with non-Gaussian deviations suppressed by slow-roll parameters as usual~\cite{Maldacena:2002vr}.

Now, in more general situations we expect a self-interaction affecting the dynamics of the isocurvature field $\psi$. Thus, instead of the Lagrangian (\ref{Lagrantian-quadratic}), we may consider the following extension:
\bea
\mathcal L (\zeta , \psi) &=& a^3 \Big[\epsilon  ( \dot \zeta - \alpha \psi ) ^2 -   \frac{\epsilon}{a^2} (\nabla \zeta)^2  \nn \\
&& +  \frac{1}{2} \dot \psi^2 - \frac{1}{2a^2} (\nabla \psi)^2  - \Delta V(\psi) \Big]. \label{Lagrantian-full-v}
\eea
Notice that the potential $\Delta V(\psi)$ is replacing the initial mass term of Eq.~(\ref{Lagrantian-quadratic}). Without any concrete knowledge of $\Delta V(\psi)$ we would expect that it could be expanded in a power series of the form:
\be
\Delta V(\psi) \simeq \frac{1}{2} \mu^2 \psi^2 + \frac{1}{6} g \, \psi^3 + \cdots . \label{hierarchical-exp-V}
\ee
However, such an expansion assumes that the coefficients $\mu^2$, $g$, etc... are such that, for amplitudes of $\psi$ characteristic of horizon crossing, the higher order terms of the expansion remain suppressed. In this work we want to explore those situations where the fluctuations $\psi$ are such that we cannot disregard the structure of $\Delta V(\psi)$ by assuming the hierarchical expansion of (\ref{hierarchical-exp-V}). To make this statement more concrete, we will consider the following specific axionlike potential:
\be
\Delta V(\psi) =  \Lambda^4 \left[ 1 - \cos \left( \frac{\psi}{f} \right) \right] , \label{potential-psi}
\ee
where $f$ is the axion decay constant. In Appendix~\ref{sec:concrete-example} we shall describe a concrete example wherein this Lagrangian emerges.

To continue, notice that the potential (\ref{potential-psi}) breaks the shift symmetry (for $\mu = 0$) of the Lagrangian (\ref{Lagrantian-quadratic}) down to a discrete symmetry:
\bea
&& \psi \to \psi' = \psi +  2 \pi n f , \\
&& \zeta \to \zeta' = \zeta +  2 \pi n f  \int^t \!\!  dt \, \alpha .
\eea
This time, $\psi$ may acquire constant solutions that minimize the sinusoidal potential. On superhorizon scales, any of these solutions will dominate the behavior of $\psi$. Just as before, $\zeta$ will be sourced by $\psi$, but this time the enhancement will happen for those values of $\psi$ that minimize the potential. This result suggests that the statistics transferred from $\psi$ to $\zeta$ will continue to be operative in this new context, but in a manner that it will be enhanced at those values in which $\psi$ coincides with a minimum of the potential, and suppressed for those values in which $\psi$ coincides with a maximum. Therefore, the structure of the potential will be necessarily inherited by the PDF of the curvature perturbation $\zeta$, which becomes non-Gaussian.

Presumably, the Lagrangian~(\ref{Lagrantian-full-v}) is the result of perturbing a more fundamental multifield theory, with a scalar field potential of the form $V = V_0 + \Delta V$ [as already mentioned, in Appendix~\ref{sec:concrete-example} we show that this is one way to derive~(\ref{Lagrantian-full-v})]. We will be interested in the regime $\Lambda^4 /3 H^2 M_{\rm Pl}^2 \ll 1$, so the potential $\Delta V$ has little to say about the background dynamics of the full system, and inflation is driven by a piece $V_0$. Then, the background equations of motion require $V_0 \sim 3 H^2 M_{\rm Pl}^2$. In addition, if $\Lambda^4 / 3M_{\rm Pl}^2H^2 \ll 1$ then the symmetry breaking is mild and, for all practical purposes, before and during horizon crossing the field $\psi$ will behave as an ultralight field. This means that at horizon crossing $\psi$ will freeze, and Eq.~(\ref{zeta-psi-N-1}) will describe how $\psi$ transfers its statistics to $\zeta$. As times passes, the nonlinearities due to $\Delta V(\psi)$ will start to become accentuated, and one expects a nonlinear contribution to Eq.~(\ref{zeta-psi-N-1}) coming from the nonlinear evolution of $\psi$ that does not freeze. To leading order in $\alpha$ we expect that any level of nonlinearity will be communicated to $\zeta$ through a non-Gaussian contribution to the $n$-point correlation functions of the form
\be
\langle \zeta^n \rangle_{\rm NG} \propto \left( \frac{\alpha}{H} \Delta N \right)^n \langle \psi^n \rangle_{\rm NG} . \label{NG-transfer}
\ee
The reason behind this guess is the following: First, in the absence of interactions between the two fields ($\alpha = 0$), the fluctuation $\psi$ will acquire a non-Gaussian distribution due to its potential $\Delta V(\psi) = \Lambda^4 [1 - \cos (\psi / f)]$. The non-Gaussian contribution to the $n$-point correlation functions were computed in Ref.~\cite{Palma:2017lww} using the in-in formalism, and are found to be given by
\bea
\langle \psi (\k_1, \tau)  \cdots \psi(\k_n , \tau) \rangle_c  =  (-1)^{n/2}  (2\pi)^3 \delta^{(3)}
\Big(\sum_j \k_j \Big) \nn \\
\times \frac{2}{3} \frac{\Lambda^4}{H^4 }  e^{ - \frac{\sigma_0^2}{2 f^2} }  \left( \frac{H^2}{2 f } \right)^{n}  \frac{  k_1^3 + \cdots + k_n^3 }{ k_1^3 \cdots k_n^3 } \Delta N, \qquad \label{n-point-psi}
\eea
where $\sigma_0^2$ is the variance of $\psi$ appearing from loop resummations. In the previous expression the subscript $c$ indicates that we are only taking into account the diagrammatic contributions due to the potential $\Delta V(\psi)$ that are fully connected [which is why there is a single overall Dirac-delta function on the right-hand side of (\ref{n-point-psi}) enforcing momentum conservation]. This set of $n$-point correlation functions are generated during horizon crossing.

Second, let us turn back on the coupling $\alpha \neq 0$. Then, because we are assuming that $\Delta V(\psi) / 3 H^3 M_{\rm Pl}^2 \ll 1$ the field $\psi$ is essentially massless and the linear relation (\ref{zeta-psi-N-1}) will remain valid on superhorizon scales, independent of the nonlinear dynamics. This implies that any non-Gaussianity gained by $\psi$ during horizon crossing will be transferred to $\zeta$ via Eq.~(\ref{NG-transfer}) after horizon crossing.  As we shall see, a detailed computation leads to
\be
\langle \zeta^n \rangle_{\rm NG}  \simeq \frac{1}{2} \left( \frac{\alpha}{H} \Delta N \right)^n \langle \psi^n \rangle_{\rm NG} , \label{n-points-zeta-psi}
\ee
where the factor $1/2$ comes from the interaction structure implied by certain nested integrals appearing in the computation of the $n$-point correlation functions using the in-in formalism.

The importance of Eqs.~(\ref{n-point-psi}) and~(\ref{n-points-zeta-psi}) is that they allow us to infer a probability distribution function for $\zeta$. This PDF is characterized by a class of non-Gaussianity that cannot be fully captured with three- or four-point correlation functions, as opposed to the case springing from the ansatz~(\ref{hierarchical-exp-V}). This PDF is found to be given by (see Sec.~\ref{subsec:PDF-derivation-from-n-points-full} for the derivation)
\bea
\rho (\zeta) = \frac{e^{- \frac{\zeta^2}{2 \sigma_{\zeta}^2}}}{\sqrt{2\pi} \sigma_{\zeta}}  \left[ 1 + A^2 \int_0^{\infty} \!\!\frac{dx}{x} \, \mathcal{K}(x)  \right. \qquad \qquad \nn \\  \left. \left(  \frac{  \sigma_{\zeta}^2 }{ f_\zeta(x)^2 }   \cos\left( \frac{\zeta}{f_\zeta(x)} \right) -  \frac{  \zeta }{ f_\zeta(x) }   \sin\left( \frac{\zeta}{f_\zeta(x)}  \right) \right)  \right] , \quad \label{distribution-0}
\eea
where $\sigma_{\zeta}$ is the variance of $\zeta$ parametrizing the Gaussian part of the distribution. In the previous expression $\mathcal K(x)$, $f_{\zeta}(x)$ and $A$ are given functions and quantities determined by parameters related to $\Delta V(\psi)$ that will be deduced in the next sections.

The main characteristic of the distribution function $\rho (\zeta)$ is that, in spite of the $x$ integral, it inherits the structure of the potential $\Delta V(\psi)$. That is, the probability of measuring $\zeta$ increases (decreases) if the field $\psi$ sourcing its amplitude is at a local minimum (maximum) of $\Delta V(\psi)$. The mechanism described here is certainly not exclusive to the potential $\Delta V(\psi)$ given in Eq.~(\ref{potential-psi}). It should be safe to suspect that any potential $\Delta V(\psi)$ with a rich structure (i.e., characterized by field distances $\Delta \psi$ smaller than $H$) will imply the existence of some level of non-Gaussianity for $\zeta$ revealing the structure of $\Delta V(\psi)$. In fact, we will prove this statement in the companion Letter~\cite{Chen:2018brw}. Thus, we see that the type of non-Gaussian departures discussed here in principle gives us nontrivial information about the landscape, offering us tomographic information about the shape of the multifield potential.

Before finishing this section, let us mention that the field $\psi$ is not expected to be a true axion as realized in QCD or string theory~\cite{Peccei:1977hh, Svrcek:2006yi} for the range of parameters that we are interested in. The reason is that large fluctuations of $\psi$ traversing many minima of the potential would destabilize the radial field fixing the value of the axion decay constant $f$~\cite{Lyth:1992tx}. For this reason, we take the potential of Eq.~(\ref{potential-psi}) to be representative of systems with potentials with a rich structure, as expected in the string landscape. See Ref.~\cite{Linde:1993xx} for a previous work that has studied the system (\ref{Lagrantian-full-v}) with an axionlike potential (for the decoupled case $\alpha = 0$) analyzing issues related to the landscape. For a recent review on the role of axions in cosmology and inflation see Ref.~\cite{Marsh:2015xka} and references therein.

In the rest of this article, we set ourselves to derive Eqs.~(\ref{n-points-zeta-psi}) and~(\ref{distribution-0}) using the in-in formalism, giving details about every step of the computations.

\section{Setting the stage}  \label{sec:stage}

One of the main technical goals of this article is to compute $n$-point correlation functions of $\zeta$ resulting from the potential $\Delta V(\psi)$ given in Eq.~(\ref{potential-psi}). To proceed, we introduce canonical fields $u$ and $v$ defined as:
\bea
u  \equiv \sqrt{2\epsilon} a \zeta,  \label{u-def} \\
v \equiv  a \psi . \label{v-def}
\eea
To simplify any computation involving $u$ and $v$ we will assume a purely de Sitter background, with $a(t)=e^{H t}$. This means that our computations will have corrections of order $\epsilon$ not accounted for. It is also convenient to introduce conformal time $\tau$ (via $d\tau = dt / a = dt e^{-Ht}$) and write
\be
a(\tau)= - 1 / H\tau , \label{a-tau}
\ee
with $- \infty <\tau < 0$. Then, putting together Eqs.~(\ref{u-def})-(\ref{a-tau}) back into the action~(\ref{Lagrantian-quadratic}), we obtain
\bea
S &=& \frac{1}{2} \int d^3 x d\tau  \Bigg[ \left( u'  + \frac{\lambda}{\tau}  v \right)^2 +\frac{2}{\tau^2} u^2    -  (\nabla u)^2  \nn \\
 &&  +  ( v' )^2 + \frac{2}{\tau^2} v^2     - (\nabla v)^2  + \lambda \frac{2}{\tau^2} u v  \nn \\
 && - 2 a^4 \Lambda^4 \bigg( 1 - \cos \Big(\frac{v}{af} \Big) \bigg) \Bigg] ,  \label{action-quadratic-u-v}
\eea
where we have introduced the dimensionless coupling
\be \label{lambda}
\lambda\equiv\frac{\sqrt{2\epsilon}\alpha }{H},
\ee
which is taken to be nonvanishing in the de Sitter limit. From Eq.~(\ref{action-quadratic-u-v}), we infer that the canonical momenta associated with $u$ and $v$ are,  respectively, given by
\bea
\Pi_u &=& u'  + \frac{\lambda}{\tau}  v , \label{can-momenta-1} \\
\Pi_v &=& v' . \label{can-momenta-2}
\eea
These momenta satisfy the equal time commutation relations, given by
\bea
\left[u(\x, \tau), \Pi_u (\y, \tau)\right] &=& i \delta^{(3)} (\x - \y) , \label{commut-rel-1} \\
\left[v(\x, \tau), \Pi_v (\y, \tau)\right]  &=& i \delta^{(3)} (\x - \y)  ,\label{commut-rel-2}
\eea
with every other commutator vanishing. From (\ref{can-momenta-1}) and (\ref{can-momenta-2}) we see that the Hamiltonian of the system is given by
\bea
H \! &=& \! \frac{1}{2} \int_x \Bigg[ \Pi_u^2 + (\nabla u)^2 - \frac{2}{\tau^2} u^2 + \Pi_v^2 + (\nabla v)^2 - \frac{2}{\tau^2} v^2 \nn \\
&& - \frac{2 \lambda}{\tau} v   \left(  \Pi_u + \frac{u}{\tau}  \right) + 2 a^4 \Lambda^4 \bigg( 1 - \cos \Big(\frac{v}{af} \Big) \bigg) \Bigg] ,
\eea
where we have introduced the notation $\int_x = \int d^3x$.

\subsection{Splitting the theory}

We may now split the Hamiltonian into three contributions as $H = H_0 + H_\lambda + H_{\Lambda}$, where $H_0$ corresponds to the free Hamiltonian of the system (obtained in the limit $\lambda = \Lambda = 0$). Notice that $H_0$ describes a system with two decoupled massless scalar perturbations
\be
H_0 =  \frac{1}{2} \int_x \left[ \Pi_u^2 + (\nabla u)^2 - \frac{2}{\tau^2} u^2 + \Pi_v^2 + (\nabla v)^2 - \frac{2}{\tau^2} v^2 \right] . \label{H-0}
\ee
On the other hand, $H_\lambda$ contains the interaction term proportional to $\lambda$,
\be
H_\lambda = - \int_x  \frac{\lambda}{\tau} v   \left(  \Pi_u + \frac{u}{\tau} \right)  , \label{H-mixing}
\ee
and $H_\Lambda$ contains the self-interactions for $v$
\be
H_\Lambda =  \int_x a^4 \Lambda^4 \bigg( 1 - \cos \Big(\frac{v}{af} \Big) \bigg) .  \label{H-Lambda}
\ee
We may now quantize the system by adopting the interacting picture framework. That is, the quantum fields $u$ and $v$ are expressed as
\bea
u (\x , \tau) &=& U^\dag (\tau)  u_I (\x , \tau)  U (\tau),  \label{u-v-U-1} \\
v (\x , \tau) &=& U^\dag (\tau)  v_I (\x , \tau)  U (\tau), \label{u-v-U-2}
\eea
where $u_I (\x , \tau)$ and $v_I (\x , \tau)$ are the interaction picture fields, which evolve as quantum fields of the free theory. Explicitly, they are given by
\bea
u_I(\x , \tau) &=& \int_k \, \hat u_I (\k , \tau) \, e^{- i \k \cdot \x }, \\
v_I(\x , \tau) &=& \int_k \, \hat v_I (\k , \tau) \,  e^{- i \k \cdot \x },
\eea
with $\int_k = (2\pi)^{-3} \int d^3 k$, and where
\bea
\hat u_I (\k , \tau) &=& u_k(\tau) a_{-}(\k) + u_k^*(\tau) a_{-}^\dag(-\k)  , \label{Fourier-u-v-1} \\
\hat v_I (\k , \tau) &=& v_k(\tau) a_{+}(\k) + v_k^*(\tau) a_{+}^\dag(-\k) . \label{Fourier-u-v-2}
\eea
Here, the pairs $a_{\pm}(\k)$ and $a_{\pm}^\dag(\k)$ correspond to creation and annihilation operators satisfying the commutation relations:
\be
\left[ a_{b} (\k) , a_{c}^{\dag} (\k') \right] = (2 \pi)^3 \delta^{(3)} (\k - \k') \delta_{bc} , \label{cre-anh-def}
\ee
with \(b,c \in \{+, -\}\). The mode functions $u_k(\tau)$ and $v_k(\tau)$ are both given by
\be
u_k(\tau) = v_k(\tau) = \frac{1}{\sqrt{2 k}} \left( 1 - \frac{i}{k \tau} \right) e^{- i k \tau} , \label{k-mode-solutions}
\ee
which corresponds to the standard expression for a massless mode on a de Sitter spacetime with Bunch-Davies initial conditions. On the other hand, $U (\tau)$ is the propagator in the interaction picture, which is given by
\be
U (\tau) = \mathcal T \exp \left\{ - i \int_{-\infty_{+}}^{\tau} \!\!\!\!\!\! d \tau' H_I (\tau')  \right\} , \label{U-propagator}
\ee
where $\mathcal T$ stands for the time ordering symbol. In a given product of operators, $\mathcal T$ instructs us to put operators evaluated at later times on the left and operators evaluated at earlier times on the right. In addition, we have $\infty_{\pm} = \infty (1 \mp i \epsilon)$, where $\epsilon$ is a small positive number introduced to select the correct interaction picture vacuum. Finally, $H_I$ in Eq.~(\ref{U-propagator}) is given by
\be
H_I = H_I^{\lambda}  + H_I^{\Lambda}   ,
\ee
where
\bea
H_I^{\lambda} &=& - \int_x  \frac{\lambda}{\tau} v_I   \left(  \Pi_u^{I} + \frac{u_I}{\tau}  \right) , \label{int-alpha} \\
H_I^{\Lambda} &=& \int_x a^4 \Lambda^4 \bigg( 1 - \cos \Big(\frac{v_I}{af} \Big) \bigg) . \label{int-Lambda}
\eea
Notice that in the previous expressions the canonical momenta $\Pi_u^{I}$ and $\Pi_v^{I}$ in the interaction picture are simply given by
\be
 \Pi_u^{I} = \frac{d}{d\tau} u_I , \qquad  \Pi_v^{I} = \frac{d}{d\tau} v_I .
 \ee
 In order to deal with $H_I^{\Lambda}$, we will consider the Taylor expansion of the cosine function as:
\be
H_I^{\Lambda} = - \frac{ \Lambda^4 }{H^4 \tau^4} \sum_{m=1}^{\infty} \int_x   \frac{(-1)^{m}}{(2m)!} \Big(\frac{H \tau }{f} v_I \Big)^{2m} .  \label{cosine-expansion}
\ee
This expansion gives us an infinite number of vertices to deal with. As we shall see in Sec.~\ref{sec:npoints}, it will be possible to resum this expansion back into an exponential contribution, leading to nonperturbative results in terms of the ratio $H/f$.

\begin{figure}[t!]
\includegraphics[scale=0.25]{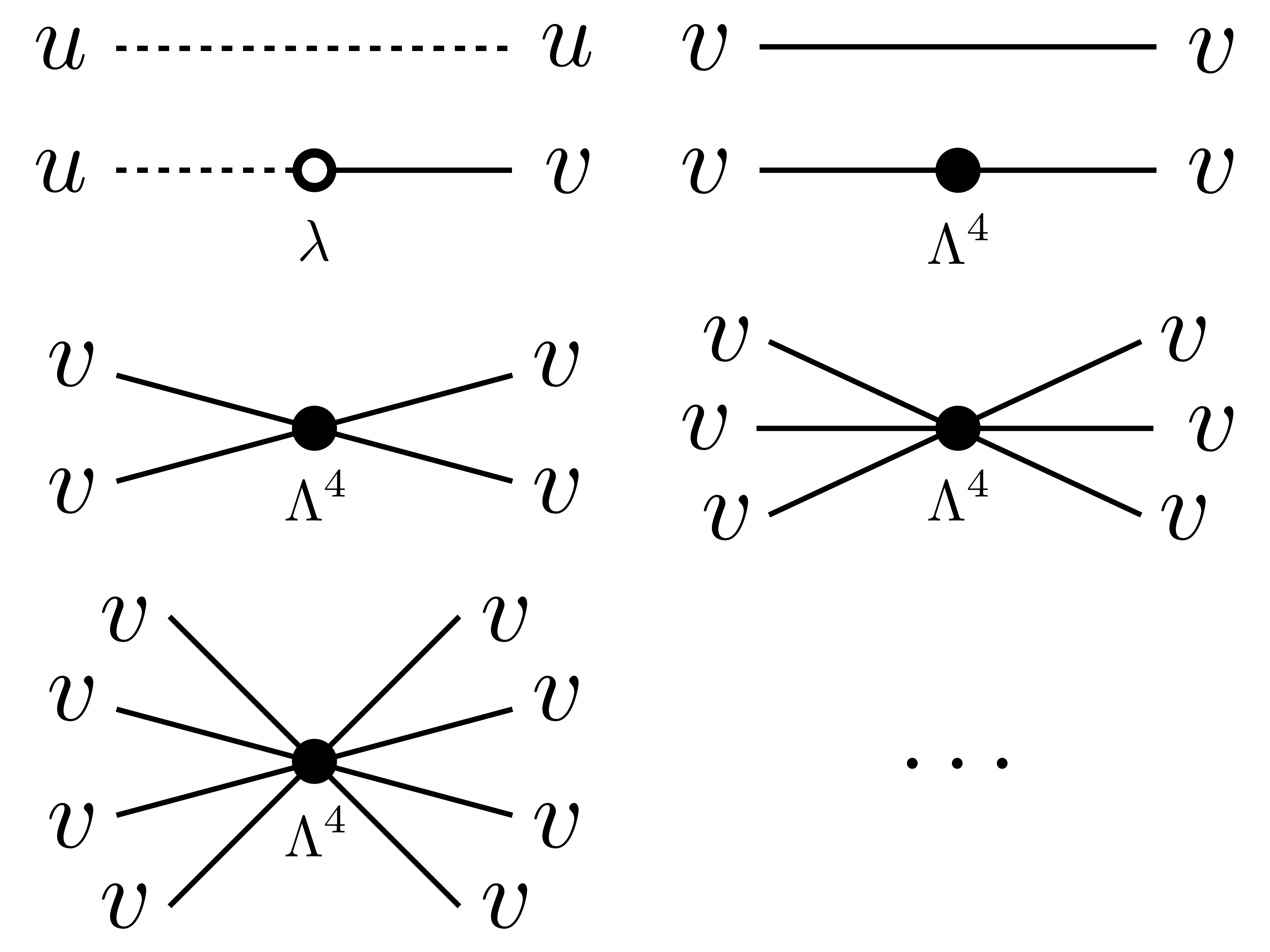}
\caption{The various diagrams of the theory. The empty circle denotes the two leg vertex offered by the Hamiltonian of Eq.~(\ref{int-alpha}). The diagrams with solid circles denote the vertices coming from the expansion (\ref{cosine-expansion}). }
\label{fig:FIG_diagrams-theory}
\end{figure}

The computation of $n$-point correlation functions of the following sections may be organized diagrammatically. These computations involve contractions of the Hamiltonians $H_I^{\lambda}$ and $H_I^{\Lambda}$ and the fields $u_I$ and $v_I$. In the present context, a contraction is the result of a commutation between creation and annihilation operators introduced in Eqs.~(\ref{Fourier-u-v-1}) and~(\ref{Fourier-u-v-2}) that results from their normal ordering.  A commutation involving the pair $a_{-}^\dag$ and $a_{-}$ is represented by a dashed line joining vertices (or external legs) labeled by the $u$ fields that contain these operators. Similarly, a commutation involving the pair $a_{+}^\dag$ and $a_{+}$ is represented by a solid line labeled by the $v$ fields that contain these operators. If the field participating in the commutation comes from $H_I^{\lambda}$, then the line joins a vertex with an empty circle. Otherwise, if the field participating in the commutation comes from $H_I^{\Lambda}$, then the line joins a vertex with a solid circle. Figure~\ref{fig:FIG_diagrams-theory} shows the various classes of diagrams appearing in the computation of $n$-point correlation functions.

\subsection{Perturbativity conditions} \label{pert-conds}

In the next sections, we compute the $n$-point correlation functions perturbatively. Given that we will resum the expansion in $H/f$ appearing in~(\ref{cosine-expansion}), we will not impose any condition on the size of $f$. On the other hand, it is worth counting with some criteria on how large $\lambda$ and $\Lambda$ can be in order to have a well-behaved perturbation theory. A naive estimation (that does not take into account renormalization) may be obtained by rewriting the Lagrangian~\eqref{Lagrantian-full-v} in a dimensionless form, weighting spacetime variables and fields by their characteristic values. In de Sitter, a characteristic length scale is given by $H^{-1}$. Moreover, the amplitude of massless scalar fields around horizon crossing is of order $H$. Thus, redefining spacetime and field variables as
\bea
t &\to& \bar t = t H ,  \\
\x &\to& \bar \x = \x H , \\
\psi &\to& \bar \psi = \frac{1}{H}\psi  , \\
\zeta &\to& \bar \zeta = \frac{\sqrt{2 \epsilon}}{H} \zeta ,
\eea
and after writing $ \bar {\mathcal L} = \mathcal L / H^4$, the Lagrangian~\eqref{Lagrantian-full-v} becomes
\bea
 \bar {\mathcal L} (\bar \zeta , \bar \psi) &=& a^3\Bigg[\frac{1}{2} \left(\partial_{\bar t} \bar \zeta - \lambda \bar \psi \right) ^2 - \frac{1}{2} \frac{(\bar{\nabla}  \bar \zeta)^2}{a^2}  \nn \\
 && +   \frac{1}{2}  ( \partial_{\bar t} \bar\psi)^2 -  \frac{1}{2}  \frac{(\bar \nabla \bar \psi)^2}{a^2}  -  \frac{1}{H^4} \Delta V (g \bar \psi) \Bigg], \nn
\eea
where $\lambda= \sqrt{2 \epsilon} \alpha / H$ is the dimensionless coupling already defined in Eq.~\eqref{lambda}. Here, derivatives are with respect to the dimensionless variables, and we have further defined the ratio $g\equiv H/f$. By asking that the dimensionless couplings remain small, we obtain the following perturbativity conditions
\be \label{perturbative-cond}
\frac{\Lambda^4}{H^4} \ll 1 \qquad \text{and} \qquad \lambda \ll 1,
\ee
for the potential~\eqref{potential-psi}. Note that the first condition is stronger than $\Lambda^4 / 3M_{\rm Pl}^2H^2\ll 1$ already required for the background not to be affected by $\psi$. Also, recall that we are not restricting the value of $g=H/f$, as the results of the next sections are nonperturbative with respect to this parameter.

As we shall see in Sec.~\ref{sec:npoints}, the loop corrections due to the resummation of the sinusoidal potential \eqref{cosine-expansion} will renormalize the bare coupling $\Lambda$. The consequence of this is that the correct perturbative parameter will turn out to be
\be
\frac{\Lambda^4_{\rm ren}}{H^4} = \frac{\Lambda^4}{H^4} e^{- \frac{\sigma_{S}^2}{2 f^2}} , \label{Lambda-ren}
\ee
where $\sigma_{S}^2$ is a short wavelength contribution to the variance of the field $\psi$ (we will compute this quantity in Sec.~\ref{subsec:n-point}).  In summary, our results will be perturbative in the couplings $\Lambda_{\rm ren}^4/H^4$ and $\lambda$, but nonperturbative in the parameter $H/f$ because \eqref{cosine-expansion} will be resummed eventually.

\section{Linear theory}  \label{sec:linearth}

Before computing the $n$-point correlation functions using the full nonlinear theory, let us have a look into the linear theory obtained in the limit $\Lambda^4 \to 0$. In the absence of nonlinear interactions, the statistics will be Gaussian, and the only meaningful quantity to compute is the power spectrum for $\zeta$. This system was investigated in Ref.~\cite{Achucarro:2016fby}, and here we show the main steps allowing one to deduce the value of the power spectrum. This result will be important later on.

\begin{figure}[t!]
\includegraphics[scale=0.22]{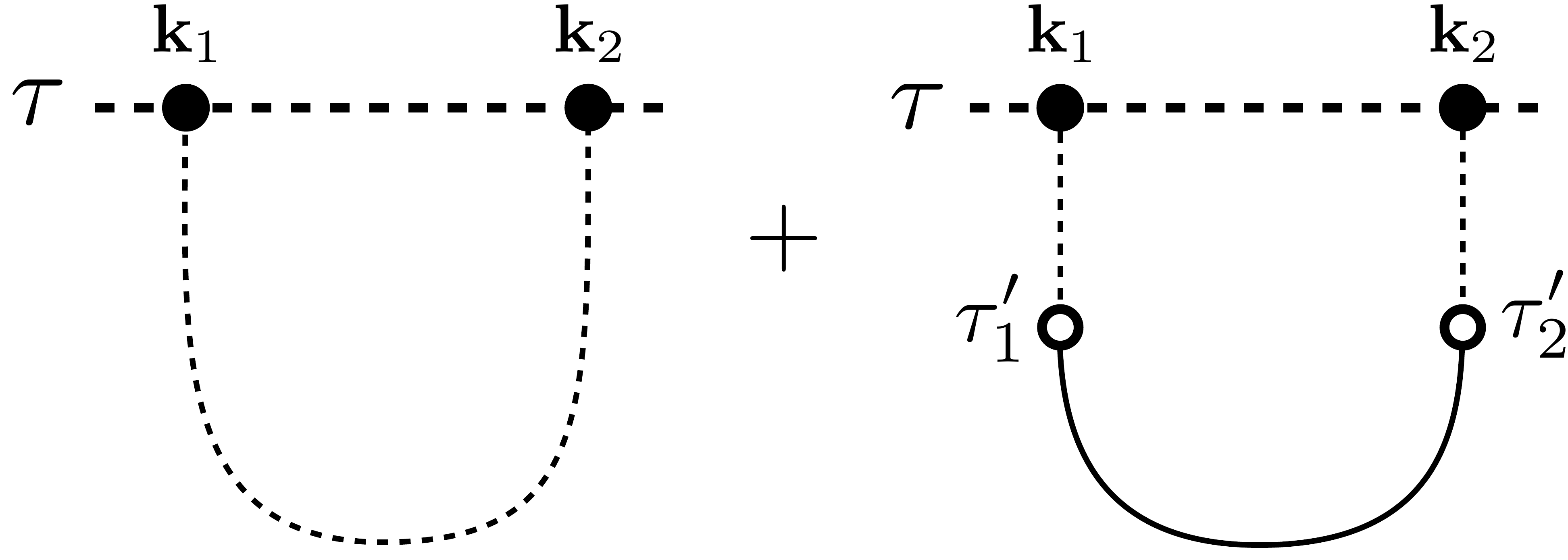}
\caption{The two diagrams contributing to the computation of the two-point function. The first diagram gives the standard power spectrum for $\zeta$, whereas the second diagram gives the correction due to the $\lambda$ derivative interaction.}
\label{fig:FIG_two-point-alpha}
\end{figure}

To start with, notice that if $\Lambda = 0$ the system consists of two canonical massless fields $u$ and $v$ coupled through the interaction Hamiltonian~(\ref{int-alpha}). The power spectrum for $\zeta$ may be obtained by computing the two-point correlation function $\langle u (\x, \tau) u(\y, \tau) \rangle$. We will perform this computation up to order $\lambda^2$. Given a fluctuation $\varphi$, we define its power spectrum $P_\varphi (k)$ as
\be
\langle  \hat \varphi (\k_1) \hat \varphi (\k_2)  \rangle = (2 \pi)^3 \delta(\k_1 + \k_2) P_\varphi (k) .
\ee
To proceed with the computation of $P_{\zeta}$, we write $u (\x , \tau) = U^{\dag} (\tau) u_I (\x, \tau) U (\tau)$, where $U(\tau)$ is given by Eq.~(\ref{U-propagator}), with $H_I$ as in (\ref{int-alpha}). Up to second order in $\lambda$ this quantity is given by
\bea
u (\x , \tau) = u_I (\x, \tau) +   i \int_{-\infty}^{\tau} \!\!\!\!\!\! d \tau' [ H_I^{\lambda} (\tau') , u_I (\x, \tau)]  \nn \\
 -  \int_{-\infty}^{\tau} \!\!\!\!\!\! d \tau' \int_{-\infty}^{\tau'} \!\!\!\!\!\! d \tau'' [ H_I^{\lambda} (\tau'') , [ H_I^{\lambda} (\tau')  , u_I (\x, \tau)]] . \label{u-uI-lambda}
\eea
The two integrals may be solved in the superhorizon limit $k |\tau| \ll 1$. The first integral is found to be given by
\bea
&& \int_{-\infty }^{\tau} \!\!\!\!\!\!\! d \tau' \left[ H_I^{\lambda} (\tau') , \hat u_I (\k , \tau) \right] =  \lim_{k \bar \tau \to - \infty} \frac{ \lambda}{(2 k)^{3/2} \tau} \bigg( \gamma
 - 2 + \ln 2
 \nn \\ &&
 - \frac{i\pi}{2}  + 2 \ln (- k \tau) - \ln ( - k \bar \tau) \bigg)  a_{+}(\k)  + {\rm H.c.} (-\k), \label{comm-linear-1}
\eea
whereas the second integral reads
\bea
&& \! \int_{-\infty }^{\tau} \!\!\!\!\!\!\! d \tau' \int_{-\infty }^{\tau'} \!\!\!\!\!\!\! d \tau'' \left[ H_I^{\lambda} (\tau'') , \left[ H_I^{\lambda} (\tau') , \hat u_I (\k , \tau) \right]  \right]  = \nn \\
&&  \lim_{k \bar \tau \to - \infty} \frac{i \lambda^2}{4 (2 k)^{3/2} \tau} \bigg( - 4 - \frac{\pi^2}{6} +
\Big[  \gamma - 2 - \frac{i \pi}{2} +  \ln 2     \nn \\ &&
+ \ln (- k \bar \tau) \Big]^2   + 2 \Big[  \gamma - 2 - \frac{i \pi}{2} +  \ln 2 + \ln (- k \bar \tau) \Big] \Big[  \gamma - 2 \nn \\ &&
- \frac{i \pi}{2} +  \ln 2 - \ln (- k \bar \tau)
+ 2 \ln (- k \tau) \Big] \bigg) a_{-}(\k) +  {\rm H.c.} (-\k) .  \nn \\ \label{comm-linear-2}
\eea
These expressions, together with~(\ref{u-uI-lambda}), allow one to compute the two-point correlation function $\langle u (\x, \tau) u(\y, \tau) \rangle$. In momentum space, one obtains
\bea \label{2pointcorrelation}
&& \langle 0 | \hat u (\k_1 , \tau) \hat u (\k_2 , \tau) | 0 \rangle = (2 \pi)^3 \delta(\k_1 + \k_2) \frac{1}{2 k^3 \tau^2} \nn \\
&&  \quad \bigg( 1 + \lambda^2 \Big[ A_1 - A_2 \ln (- k \tau) + \ln^2(- k \tau ) \Big]  \bigg) , \label{power-linear}
\eea
where $A_1$ and $A_2$ are numbers given by
\bea
A_1 &=& - \frac{\pi^2}{6}  +  (3 - \ln 2) (1 - \ln 2) -  \gamma (4 - \gamma - 2 \ln 2)  , \qquad \\
A_2 &=&  4 -2 \gamma - 2 \ln 2 .
\eea
Their numerical values are $A_1 \simeq - 2.11$ and $A_2 \simeq 1.46$. Note that in putting together~(\ref{u-uI-lambda})-(\ref{comm-linear-2}) to compute the two-point correlation function, the divergent logarithms $\ln (- k \bar \tau)$ cancel out. The computation of the two-point correlation function of Eq.~(\ref{power-linear}) may be thought of as the result of the diagrammatic expansion of Fig.~\ref{fig:FIG_two-point-alpha}. The zeroth order contribution corresponds to the first diagram, whereas the contribution of order $\lambda^2$ corresponds to the second diagram, where the two external legs are mediated by a $v$ propagator.

Now, horizon crossing happens when $k |\tau| \simeq 1$. Thus, the number of $e$-folds after horizon crossing is given by
\be
\Delta N = - \ln (- k \tau) . \label{Delta-N-2-point}
\ee
It may be seen that after several $e$-folds the contribution to the power spectrum~(\ref{power-linear}) quadratic in $\Delta N$ dominates, and we obtain
\bea
&& \langle 0 | \hat u (\k_1 , \tau) \hat u (\k_2 , \tau) | 0 \rangle = (2 \pi)^3 \delta(\k_1 + \k_2) \frac{\lambda^2}{2 k^3 \tau^2} \Delta N^2  . \qquad \label{two-point-lambda-N}
\eea
For this to happen, we need $\lambda^2 \Delta N^2 \gtrsim 1$. The power spectrum for $v$ may be found through a similar computation, which gives
\bea
&& \langle 0 | \hat v (\k_1 , \tau) \hat v (\k_2 , \tau) | 0 \rangle = (2 \pi)^3 \delta(\k_1 + \k_2) \frac{1}{2 k^3 \tau^2}  ,
\eea
valid up to order $\mathcal O (\lambda^4)$. Combining these two results, we then derive the following relation between the two power spectra
\be
P_\zeta (k) =   \frac{\lambda^2}{2 \epsilon} \Delta N^2 P_\psi (k) , \label{power-zeta-psi}
\ee
where $P_\psi (k)$ is found to be given by
\be
P_\psi (k) = \frac{H^2}{2 k^3 } .
\ee
This result is consistent with the behavior shown in Eq.~(\ref{zeta-psi-N-1}) based on symmetry arguments. It shows that the power spectrum for $\zeta$ is proportional to the power spectrum of $\psi$, with a factor that grows with the number of $e$-folds.

Let us briefly comment on the validity of the result shown in Eq.~(\ref{two-point-lambda-N}). Our perturbative method consisted of separating the theory between a zeroth order Hamiltonian $H_0$ [given in Eq.~(\ref{H-0})] and an interaction Hamiltonian (given in Eq.~(\ref{int-alpha})), proportional to $\lambda$. On the one hand, we have argued that, in our final result for the power spectrum (\ref{power-linear}), we are allowed to retain as the dominant piece the term proportional to $\lambda^2$. On the other hand, notice that our perturbative method is valid as long as $\lambda^2 \ll 1$. These two statements are not in contradiction: the computation admits a cumulative effect that grows with the number of $e$-folds as $\lambda^2 \Delta N^2$, which may be larger than $1$ (after $\Delta N \simeq 60$). This effect was discussed in detail in Ref.~\cite{Achucarro:2016fby}, and it will play an important role in Sec.~\ref{sec:tom-NG}.

Also, the example of the derivative coupling we used here has a special property that, at superhorizon scales, the linear equation of motion for the isocurvaton field $\psi$ has no source term from the curvature mode $\zeta$. Therefore, $\psi$ does not grow once it exits the horizon, while $\zeta$ does. This means that, if we were to solve the coupled linear equation iteratively to all orders, the enhancement factor from $\Delta N$ would stay at the order $\Delta N^2$. Therefore, the requirement of the perturbation theory is only that $\lambda \ll 1$, and $\lambda \Delta N$ can be greater than 1.

Last but not least, even if the two conditions $\lambda^2 \Delta N^2 \gtrsim 1$ and $\lambda^2 \ll 1$ may seem to be fine-tuned, the condition $\lambda^2 \ll 1$ has only been adopted in order to be able to perform analytic computations. The requirement $\lambda^2 \Delta N^2 \gtrsim 1$ is valid independent of the perturbativity condition $\lambda^2 \ll 1$, and can already be inferred from the symmetry arguments around Eq.~(\ref{Pzeta-Ppsi}). In principle, Eq.~(\ref{Pzeta-Ppsi}) [or Eq.~(\ref{power-zeta-psi})] should be valid independently of the value of $\lambda$.

\section{Computation of correlation functions}  \label{sec:npoints}

\begin{figure*}[t!]
\onecolumngrid
\includegraphics[scale=0.12]{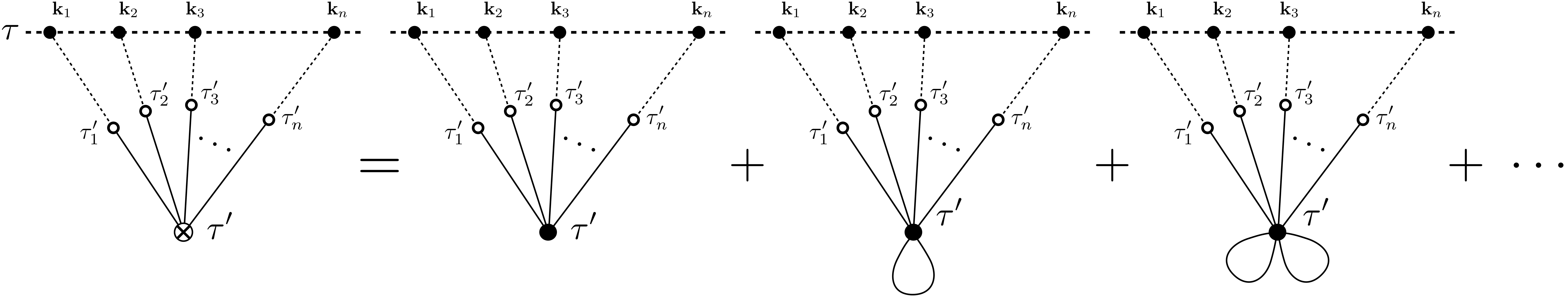}
\caption{All the connected diagrams contributing to $\tilde G^{(n)}$ at order $\Lambda^4$. The $\otimes$ vertex represents the resummation of all the loop contributions coming from the expansion of the cosine function shown in Eq.~(\ref{cosine-expansion}). In other words, for a given number of legs, the $\otimes$ vertex contains all the relevant effects due to the cosine. Because of the combinatorial factors of each diagram, and given that there are no external momenta running through the loops, the resummation reduces to a constant factor given by $\exp({- \sigma^2 / 2 f^2})$.}
\label{fig:FIG_diagram_res}
\twocolumngrid
\end{figure*}

In this section, we describe how to compute the $n$-point correlation functions of $\zeta$ at the end of inflation (details of this computation are shown in Appendix~\ref{sec:details}). The quantity of interest corresponds to
\be
\tilde G^{(n)}_{\zeta} (\tau, \k_1 , \cdots , \k_n)  \equiv \langle \zeta (\k_1, \tau)  \cdots \zeta(\k_n , \tau) \rangle,
\ee
which, in terms of the canonically normalized fields introduced in Sec.~\ref{sec:stage}, is given by
\be
\tilde G^{(n)}_{\zeta}  = \left(  \frac{H |\tau| }{\sqrt{2 \epsilon}} \right)^n \langle \hat u (\k_1, \tau)  \cdots \hat  u(\k_n , \tau) \rangle . \label{n-point-u-1}
\ee
Our main goal is to obtain an expression for this function up to order $\Lambda^4$. Given that the interaction Hamiltonian~(\ref{int-Lambda}) determined by the potential $\Delta V(\psi)$ does not depend on $u_I$, the $n$-point correlation functions will acquire a dependence on $\Lambda^4$ only through the mixing Hamiltonian~(\ref{int-alpha}) involving the coupling $\lambda$. This means that the fully connected contribution to (\ref{n-point-u-1}) will necessarily involve at least one factor $\lambda$ per field $u(\x_n , \tau)$. In other words, the lowest order contribution to (\ref{n-point-u-1}) represented by fully connected diagrams, will be of order $\Lambda^4 \lambda^n$. The diagrammatic representation of this computation in momentum space is shown in Fig.~\ref{fig:FIG_diagram_res}. The $\otimes$ vertex denotes the exact vertex, up to order $\Lambda^4$, connecting the $n$ external legs participating in (\ref{n-point-u-1}). Because the expansion~(\ref{cosine-expansion}) contains an infinite number of vertices (with an even number of legs), the exact vertex consists of the sum of an infinite number of diagrams involving loops that start and finish on the same vertices. Due to overall momentum conservation, these loops do not carry external momenta.

To proceed with the computation of $\tilde G_\zeta^{(n)}$, we start by recalling that a given $u (\k_j , \tau)$ entering the $n$-point function $\langle \hat u (\k_1, \tau)  \cdots \hat  u(\k_n , \tau) \rangle$ of Eq.~(\ref{n-point-u-1}) has the form  $u (\k_j , \tau) = U^{\dag} (\tau) u_I (\k_j, \tau) U (\tau) $.  Let us for a moment disregard the $\epsilon$ prescription determining the integration limits $\infty_{\pm}$. Then, by expanding the propagator $U (\tau)$ in this expression, we obtain
\bea
u (\k_j , \tau) = u_I (\k_j , \tau) +   i \int_{-\infty}^{\tau} \!\!\!\!\!\! d \tau' [ H_I (\tau') , u_I (\k_j , \tau)]  \qquad \qquad  \nn \\
 -  \int_{-\infty}^{\tau} \!\!\!\!\!\! d \tau' \int_{-\infty}^{\tau'} \!\!\!\!\!\! d \tau'' [ H_I (\tau'') , [ H_I (\tau' ) , u_I (\k_j , \tau)]]  + \cdots . \qquad
\eea
We only need to keep terms up to order $\Lambda^4$. Then, because $H_I^{\Lambda}$ commutes with $u_I$, but not with $v_I$, the previous equation may be further reduced to
\bea
u (\k_j , \tau)  = u_I (\k_j , \tau) +   i \int_{-\infty}^{\tau} \!\!\!\!\!\! d  \tau_j   [ H_I^{\lambda} (\tau_j) , u_I (\k_j , \tau) ]  \nn \\
 -  \int_{-\infty}^{\tau} \!\!\!\!\!\! d \tau_j \int_{-\infty}^{\tau_j} \!\!\!\!\!\! d \tau' [ H_I^{\Lambda} (\tau') , [ H_I^{\lambda} (\tau_j)  , u_I (\k_j , \tau) ]] \nn \\
 +\sum_{p=1}^{n-1} i^{p+2} \int_{-\infty}^{\tau} \!\!\!\!\!\! d \tau_j \int_{-\infty}^{\tau_j} \!\!\!\!\!\! d \tau' \int_{-\infty}^{\tau'} \!\!\!\!\!\! d \tau'_1 \cdots   \int_{-\infty}^{\tau'_{p-1}} \!\!\!\!\!\! d \tau'_p  \nn \\
  \left[
  H^{\lambda}_I (\tau'_{p}) , \cdots \left[
  H^{\lambda}_I (\tau'_1) , \left[ H^{\Lambda}_I (\tau') , \left[ H^{\lambda}_I  (\tau_j)  , u_I (\k_j , \tau) \right] \right]
   \right] \cdots
    \right]
 \nn \\
 + \cdots, \qquad  \label{u-u_I}
\eea
where the ellipses of the last line denote terms that will not contribute to the piece that we want to compute. For instance, by inserting another \(H^{\lambda}_I\) Hamiltonian (through a commutator) between \(\tau'\) and \(\tau_j\) we would be computing a correction to the \(\zeta\) propagator and not to the fully connected part of order \(\Lambda^4 \lambda^n\).

Now, Eq.~(\ref{u-u_I}) tells us that the structure of $u (\k_j , \tau)$ in terms of creation and annihilation operators $a_{\pm}$ is of the following form:
\bea
u &\sim& a_{-} + \lambda a_{+} + \Lambda^4 \lambda \sum_{m} a_{+}^{2m-1}  + \Lambda^4 \lambda^2  a_{-} \sum_{m} a_{+}^{2m-2}   \nn \\
&& + \cdots + \Lambda^4 \lambda^n  a_{-}^{n-1} \sum_{m} a_{+}^{2m-n} + \cdots  .  \label{structure-u}
\eea
The computation of $n$-point correlation functions requires us to plug this form of $u$ back into (\ref{n-point-u-1}) and perform every possible contraction between creation and annihilation operators of the various terms appearing in (\ref{structure-u}). The final result that we are pursuing is an expression containing only terms of order $\Lambda^4 \lambda^n$, and thus many of the contractions correspond to loops involving pairs of $a_{+}$ operators. The diagrammatic expansion of this computation is shown in Fig.~\ref{fig:FIG_diagram_res}. Since we are computing the fully connected contribution to (\ref{structure-u}), in every contraction involving $a_{+}$ operators, at least one of them must come from a term of order $\Lambda^4$ in (\ref{structure-u}).

The details of this computation are shown in Appendix~\ref{sec:details}. The final result is found to be given by
\bea
&& \tilde G_{\zeta}^{(n)} (\tau, \k_1 , \cdots , \k_n) =  (-1)^{n/2} (2 \pi)^3 \delta^{(3)} \Big(\sum_j \k_j \Big) \frac{\Lambda^4}{3H^4} \nn \\
&& \quad e^{- \frac{\sigma_0^2}{2 f^2}} \left( \frac{\lambda H^2 \Delta N}{2 f \sqrt{2 \epsilon}} \right)^n  \frac{k_1^3 + \cdots + k_n^3}{k_1^3 \cdots k_n^3} \Delta N , \label{n-point-final-res-zeta}
\eea
for even $n$, as it vanishes for odd $n$ because the potential is even under $\psi \to -\psi$. Here $\sigma_0$ is the variance of $\psi$, defined through the relation
\be
\sigma_0^2 \equiv \langle \psi^2 \rangle = H^2 \tau^2 \int_k u_{k} (\tau) u_{k}^* (\tau)  . \label{sigma_0-def}
\ee
The factor $\exp(- \sigma_0^2 / 2 f^2)$ of Eq.~(\ref{n-point-final-res-zeta}) appears as the consequence of the resummation shown in Fig.~\ref{fig:FIG_diagram_res}. This result may be compared to that shown in Eq.~(\ref{n-point-psi}) and obtained in Ref.~\cite{Palma:2017lww}. This comparison proves the result quoted in Eq.~(\ref{n-points-zeta-psi}) of the Introduction.

In Eq.~(\ref{n-point-final-res-zeta}), $\Delta N$ corresponds to the number of $e$-folds elapsed since a reference time $\tau_0$ around which the set of modes $k_l$ crossed the horizon ($k_l \tau_0 \sim 1$)
\be
\Delta N = \ln \left( \frac{\tau_0}{\tau} \right) .
\ee
This definition coincides with that of Eq.~(\ref{Delta-N-2-point}) in the case of the two-point function.
Notice that the shape in momentum space of the non-Gaussianity parametrized by these $n$-point correlation functions is of the local type. In obtaining this result we have assumed that the condition $ \Delta N \gg 1$ holds. If instead one has a relatively small $\Delta N $, then other terms that were neglected might need to be included back. Given that our perturbativity conditions demand $\lambda \ll 1$, this is in agreement with the power spectrum shown in (\ref{power-zeta-psi}), valid for $\lambda^2 \Delta N^2 \gtrsim 1$.

\subsection{About horizon exit} \label{sec:npointshorizon}

Here let us make a remark on the value of $\Delta N$. Since the $\psi$ field in our model example is exactly massless, $\Delta N$ should be evaluated from when modes exit the horizon until the end of inflation. Because longer modes exit the horizon earlier, $\Delta N$ is not exactly a constant and has a logarithmic dependence on the magnitude of relevant momenta. Here we ignore this weak momentum dependence and approximate $\Delta N \sim 60$ as a constant. On the other hand, for our purpose we do not have to regard the mass of the scalar field $\psi$ to be exactly zero (but still require $\mu \ll H$, so our model conditions are satisfied). Such a field would have decayed before it stays at the superhorizon for the entire 60 $e$-folds, as it happens in the lower mass range case of the quasi-single-field inflation models \cite{Chen:2009we,Chen:2009zp}. In this case, all modes of $\psi$ will stay for the same amount of $e$-folds, $\Delta N$, after the horizon exit and before the decay, and $1<\Delta N<60$ is now exactly a constant.

\subsection{Regularization: IR and UV cutoffs}

Before we finish this section, let us briefly come back to Eq.~(\ref{sigma_0-def}). If we replace the mode solutions $u_k(\tau)$ of Eq.~(\ref{k-mode-solutions}) back into~(\ref{sigma_0-def}), and define the dimensionless integration variable $q = k |\tau|$, we obtain
\be
\sigma_0^2 = \frac{H^2}{4 \pi^2} \int \!  dq  \left( q + \frac{1}{q} \right) . \label{variance-0}
\ee
Observe that $\sigma_0$ is independent of time $\tau$. However, it contains divergences coming from the integration limits $q\to0$ and $q \to +\infty$.
We may therefore introduce infrared and ultraviolet cutoff scales $q_{\rm IR}$ and  $q_{\rm UV}$, respectively, and obtain
\be
\sigma_0^2 = \frac{H^2}{4 \pi^2}   \left( \frac{1}{2} (q_{\rm UV}^2 - q_{\rm IR}^2) + \ln (q_{\rm UV}/q_{\rm IR}) \right).
\ee
Notice that the variable $q = k |\tau| = k / a H$ is the physical momenta per unit of the Hubble scale.
The UV cutoff refers to a scale that is deep inside the horizon and is the scale of new physics and the limit of low energy effective theory. This cutoff contributes to the renormalization of the coupling $\Lambda^4$ as in what happens in flat spacetime. The logarithmic IR divergence is due to the random walk of the massless field in the dS space. Actual observations do not have access to all the scales, and so the IR cutoff should be set by the size of the observable Universe. We will come back to this issue in Sec.~\ref{subsec:n-point}, where we consider the need of defining the variance of modes available to cosmological observers.

\section{Tomographic non-Gaussianity} \label{sec:tom-NG}

The expression for the $n$-point correlation functions given by (\ref{n-point-final-res-zeta}) may be Fourier transformed back into coordinate space as
\be
G_{\zeta}^{(n)} (\tau, \x_1 , \cdots ) = \int_{k_1} \!\!\!  \cdots \int_{k_n} \!\!\! e^{- i \sum_j \k_j \cdot \x_j}  \tilde G_{\zeta}^{(n)} (\tau, \k_1 , \cdots ) . \label{G-xi}
\ee
This expression may be used to deduce the probability distribution function of measuring an amplitude $\zeta({\bf x})$ at a given position ${\bf x}$. To this end, we need to compute the moments $\langle \zeta^n \rangle$ that are given by evaluating all the coordinates in (\ref{G-xi}) at a common value $\x$
\be
 \langle \zeta^n \rangle_c \equiv G_{\zeta}^{(n)} (\tau, \x , \cdots , \x) , \label{zeta-n-G}
\ee
where the subscript $c$ denotes that $\langle \zeta^n \rangle_c$ comes from fully connected diagrams, hence it is proportional to a Dirac-delta that conserves momentum. Due to this momentum conservation, $ \langle \zeta^n \rangle_c$ defined in Eq.~(\ref{zeta-n-G}) is independent of $\x$. In the following subsections, we first obtain a concrete expression for the $n$-point functions of Eq.~(\ref{zeta-n-G}) valid for long wavelength modes, and then we proceed to derive the PDF from where these $n$-point functions are computed.

\subsection{$n$-point functions for long wavelength modes} \label{subsec:n-point}

The quantity $\tilde G_{\zeta}^{(n)} (\tau, \k_1 , \cdots , \k_n)$ of Eq.~(\ref{n-point-final-res-zeta}) only shows the leading IR contribution to the full $n$-point correlation function. For the same reason, in (\ref{G-xi}) we cannot integrate along the entire momentum space, and we are forced to introduce a cutoff momentum $k_L$. This is not a technical limitation, but all the contrary. We are interested in making predictions of inflation valid for superhorizon perturbations (that will later on reenter the horizon after inflation), and so we want to compute correlation functions of long wavelength $\zeta$ modes. This is normally done by introducing window functions selecting the relevant scales for the computation of correlation functions in coordinates space. For simplicity, here we consider a window function with a hard cutoff $k_L$. With this purpose in mind, we introduce the cutoff in terms of physical momentum $q_{\rm phys} \equiv k |\tau|$ (per unit of $H$) instead of comoving momentum $k$. That is, we choose a hard cutoff momentum $q_L$ and split the curvature perturbation as
\be
\zeta = \zeta_S + \zeta_L ,  \label{scale-division}
\ee
where $\zeta_L$ only includes modes of wavelengths larger than some fixed value $2 \pi / q_L$. Horizon crossing happens at $q_{\rm phys} = 1$, and so we must impose 
\be
q_L \leq 1 .
\ee
In other words $\zeta_L$ contains superhorizon contributions (at the end of inflation) between the physical cutoff scales $q_L$ and $q_{\rm IR}$. Explicitly, $\zeta_L$ is given by
\be
\zeta_L (\x, \tau) = \int_{k<k_L} \zeta(\k,\tau) e^{- i\k \cdot \x} ,
\ee
where $k_L = q_L / |\tau|$. Thus, we will compute a more restricted version of (\ref{zeta-n-G}) given by
\be
 \langle \zeta_L^n \rangle_c = G_{\zeta,L}^{(n)} (\tau, \x , \cdots , \x) , \label{zeta-n-G-L}
\ee
where $G_{\zeta,L}^{(n)} (\tau, \x_1 , \cdots , \x_n)$ reads as in (\ref{G-xi}), but now with the momenta integrated up to $k_L$. Explicitly, we have
\bea \label{n-point-corr-space-cut}
&& G_{\zeta,L}^{(n)} (\tau, \x_1 , \cdots , \x_n) =  (-1)^{n/2} (2 \pi)^3 \frac{\Lambda^4}{3H^4} \nn \\
&& \quad  \int_{k_1 < k_L}  \cdots \int_{k_n < k_L} \delta^{(3)} \Big(\sum_j \k_j \Big) e^{- i \sum_j \k_j \cdot \x_j} \nn \\
&& \quad e^{- \frac{\sigma_0^2}{2 f^2}} \left( \frac{\lambda H^2}{2 f \sqrt{2 \epsilon}} \Delta N \right)^n  \frac{k_1^3 + \cdots + k_n^3}{k_1^3 \cdots k_n^3} \Delta N . \quad \label{n-point-final-res-coord}
\eea
The division of scales (\ref{scale-division}) forces us to split $\sigma_0^2 = \langle \psi^2 \rangle$, introduced in Eq.~(\ref{variance-0}), into short and long wavelength contributions as $\sigma_0^2 = \sigma_S^2 + \sigma_L^2$, in such a way that $\sigma_S^2$ and  $\sigma_L^2$ receive contributions larger and smaller than $k_L$ respectively. From Eq.~(\ref{variance-0}) we see that $\sigma_S^2$ and $\sigma_L^2$ are given by
\bea
\sigma_S^2 &=&\frac{H^2}{4 \pi^2}   \left( \frac{1}{2} (q_{\rm UV}^2 - q_{L}^2) + \ln (q_{\rm UV}/q_{L}) \right) \nn \\ 
&\simeq& \frac{H^2}{8 \pi^2}  q_{\rm UV}^2  , \label{sigma-S-cut-off} \\
\sigma_L^2 &=&  \frac{H^2}{4 \pi^2}   \left( \frac{1}{2} (q_{L}^2 - q_{\rm IR}^2) + \ln (q_{L}/q_{\rm IR}) \right) \nn \\
 & \simeq& \frac{H^2}{4 \pi^2} \ln \xi  , \label{sigma-cut-off}
\eea
where we have introduced the ratio
\be
\xi = \frac{q_L}{q_{\rm IR}} = \frac{k_L} {k_{\rm IR}}, \label{beta-def}
\ee
which is a measure of the range of scales spanned by the long mode contributions $\zeta_L$.  The logarithmic dependence of (\ref{sigma-cut-off}) suggests that $\sigma_L^2 \simeq H^2$. For instance, if we take $q_L = 1$, then $\xi$ corresponds to the ratio between the largest wavelength and the Hubble radius at the end of inflation. Then $\ln \xi \simeq 60$, and one obtains $\sigma_L^2 \simeq H^2$. Notice, however, that, in general, $\xi$ parametrizes the window function selecting the scales, hence its value should be determined by the range of momenta available to cosmological observations. In the particular case of the CMB, this ratio is approximately given by $\ln \xi \sim 8$.

Next, to obtain an expression for $\langle \zeta^n_L \rangle_c $, we evaluate the arguments of (\ref{n-point-final-res-coord}) at a single coordinate value $\x$. Because of momentum conservation, the argument of the exponential vanishes, and we are left with the following expression
\be
 \langle \zeta^n_L \rangle_c =  (-1)^{n/2}  g_n A^2  e^{- \frac{\sigma_L^2}{2 f^2}} \left[ \frac{\lambda \sigma_L^2}{f \sqrt{ 2 \epsilon}  } \Delta N   \right]^n  , \label{n-point-final-res-coord-sol1}
\ee
where
\be
A^2 \equiv \frac{\Delta N}{6 \sigma_L^2 } \frac{\Lambda^4_{\rm ren}}{H^2} . \label{A-def}
\ee
Here, $\Lambda^4_{\rm ren} = e^{- \sigma_{S}^2 / 2 f^2} \Lambda^4$ is the renormalized coupling  introduced in Eq.~(\ref{Lambda-ren}) resulting from the loop resummation introduced in Sec.~\ref{sec:npoints}. Because this resummation is always induced by $\Lambda$, the combination $\Lambda_{\rm ren}^4 $ will be present at all orders in perturbation theory (disregarding higher order loop corrections carrying external momenta that start appearing at order $\Lambda^8$).  In Ref.~\cite{Chen:2018brw} we discuss the renormalization of $\Delta V(\psi)$ more generally, paying special attention to the running of the parameters defining $\Delta V(\psi)$ in order to make observables independent of the cutoff scales. To continue, the coefficient $g_n$ in (\ref{n-point-final-res-coord-sol1}) is defined as
\be
g_n \equiv \frac{  (2 \pi)^3 }{( 2 \sigma_L^2 / H^2 )^{n-1}} I_n, \label{gn-def}
\ee
for even $n$ and zero otherwise because $\tilde{G}_{\zeta}^{(n)}$ vanishes if $n$ is odd. Here $I_n$ corresponds to the following integral:
\be
I_n \equiv   \int_{k_1 < k_L}  \!\!\!\!\!\!\!\!   \cdots \int_{k_n < k_L} \!\!\!\!\!\!\! \delta^{(3)} \Big(\sum_j \k_j \Big) \frac{k_1^3 + \cdots + k_n^3}{k_1^3 \cdots k_{n}^3} .  \label{full-I-n}
\ee
Equation (\ref{n-point-final-res-coord-sol1}) is written in terms of the variance $\sigma_L^2$ associated with the probability distribution function of $\psi$. It will be more useful to write $\langle \zeta^n_L \rangle_c$ in terms of the variance $\sigma_{\zeta}^2$ instead of $\sigma_L^2$. Recall that in Sec.~\ref{sec:linearth} we derived the power spectrum of $\zeta$ in terms of the power spectrum of $\psi$, given in Eq.~(\ref{power-zeta-psi}). When $\lambda^2 \Delta N^2 \gtrsim 1$, this result implies
\be
\sigma_\zeta^2 = \sigma_L^2 \frac{\lambda^2}{2 \epsilon} \Delta N^2 . \label{sigma-zeta}
\ee
Then, by defining $f_\zeta$ as
\be
f_{\zeta} \equiv f \frac{\sigma_\zeta}{\sigma_L} = f  \frac{\lambda}{\sqrt{2 \epsilon}} \Delta N  ,  \label{f-sigma-zeta}
\ee
it is direct to find
\bea
&& \langle \zeta^n_L \rangle_c =  (-1)^{n/2} g_n A^2 e^{- \frac{\sigma_\zeta^2}{2 f_\zeta^2}} \left[ \frac{\sigma_\zeta^2}{f_\zeta}   \right]^n  .  \label{zeta-n-1}
\eea
This is the general form of the $n$-point correlation function that we need in order to reconstruct the tomographic PDF for $\zeta$. It is important to emphasize here that we can also consider the regime $\lambda^2 \Delta N^2 < 1$, where one finds $\sigma_\zeta^2 = \sigma_L^2 / 2 \epsilon$. This case would give us a slightly different expression for (\ref{zeta-n-1}) but would not change the form of the reconstructed PDF (except for the way in which some parameters appear). For simplicity, we stick to the regime $\lambda^2 \Delta N^2 \gtrsim 1$.

\subsection{Dependence of the $n$-point functions on the cutoff scales}

The integral $I_n$ determining the form of the factor $g_n$ [through Eq.~(\ref{gn-def})] is a function of the order $n$ and the ratio $\xi$ introduced in Eq.~(\ref{beta-def}). Indeed, in Appendix~\ref{app_about_I_n}, we show that $I_n$ can be written in terms of a single integration variable as
\be
I_n (\xi) = \frac{n }{(2 \pi^2)^{n+1}} \int_0^{\infty} \!\! \frac{dx}{x} G(\xi,x) \left[ F(\xi , x) \right]^{n-1}   ,  \quad \label{full-I_n} 
\ee
where $G(\xi , x)$ and $F(\xi , x)$ are given by
\bea
G(\xi , x) &=&[ \sin (x) - x \cos (x) - \sin (x / \xi)  \nn \\ 
&&  + (x / \xi) \cos (x / \xi) ] , \qquad \\
F(\xi , x) &=& {\rm Ci} (x) - \frac{\sin (x)}{x}  - {\rm Ci} ( x / \xi ) + \frac{\sin ( x / \xi )}{ x / \xi} . \qquad
\eea
Here $ {\rm Ci} (x)$ is the cosine integral function. In Appendix~\ref{app_about_I_n}, we also show that in the formal limit $\xi \to \infty$, the integral asymptotes to a simple function $I_n^{0}  (\xi)$ given by
\be
 I_n^{0}  (\xi) \equiv \frac{n \pi}{2 (2 \pi^2)^{n+1}} ( \ln \xi )^{n-1} . \label{asympt-I_n}
\ee
Then, given that $\ln \xi = 4 \pi^2 \sigma^2_L / H^2$ [which, can be seen from Eq.~(\ref{sigma-cut-off})], one obtains $g_n = n$, implying a very simple general expression for the $n$-point functions $\langle \zeta^n_L \rangle_c$. However, given that $k_L$ is at most the horizon exit scale, and that $k_{\rm IR}$ is the largest scale available to present observers, we have the bound
\be
\ln \xi \leq 60,
\ee 
which implies that $\ln \xi$ is too small to allow us to take $I_n$ as $I_n^{0}$. The reason for this is that, with $\ln \xi \simeq 60$, the correction $\Delta I_n = I_n - I_n^{0}$ is already one-tenth of $I_n^0$ for $n \sim 35$. This in turn implies that the PDF derived with $ I_n^{0}$ starts to deviate significantly from the one derived with $I_n$ when $f$ is smaller than $\sigma_L \sim H$, which is precisely the interesting region of parameters that we wish to explore. A proof of this statement is given in Appendix~\ref{app_about_I_n}.

In what follows, we devote ourselves to reconstruct the PDF out of the $n$-point function $\langle \zeta^n_L \rangle_c$ given in Eq.~(\ref{n-point-final-res-coord-sol1}). We will first do this in Sec.~\ref{subsec:PDF-derivation-from-n-points} for the case in which $I_n$ is taken to be as $I_n^{0}  (\xi)$ shown in Eq.~(\ref{asympt-I_n}). Then, in Sec.~\ref{subsec:PDF-derivation-from-n-points-full},  we will show how to obtain the PDF for the full expression for $I_n (\xi)$ shown in Eq.~(\ref{full-I_n}). Before deriving these two PDFs we describe the general idea behind its reconstruction.

\subsection{PDF reconstruction: general idea}

Recall that we have focused our interest on the computation of the fully connected contributions to the $n$-point correlation functions. Had we focused instead on the full $n$-point correlation functions, including disconnected diagrams, we would have arrived at the more general expression
\be
\langle \zeta_L^n \rangle = \sum_{m=0}^{n/2} \frac{n!}{m! (n-2m)! 2^m} \sigma_{\zeta}^{2m} \langle \zeta^{n-2m}_L \rangle_c . \label{full-n-point}
\ee
Here, the factors $\sigma_{\zeta}^{2m}$ come from propagators connecting pairs of external lines. The combinatorial factor ${n!} / {m! (n-2m)! 2^m}$ consists of the total number of ways to connect the $n$ external legs in such a way that $2m$ of them are connected by propagators, and the rest  are connected to the $\Lambda^4$ vertex.

The probability distribution function $\rho(\zeta)$ that we are searching for must be such that
\be
\langle \zeta_L^n \rangle = \int \! d\zeta \, \rho(\zeta) \, \zeta^n .
\ee
To find $\rho(\zeta)$, it is useful to notice that the term $m=n/2$ in Eq.~(\ref{full-n-point}) is given by
\be
\langle \zeta_L^n \rangle_{G} = \frac{n!}{(n/2)! 2^{n/2}} \sigma_{\zeta}^{n} , \label{Gaussian-n-points}
\ee
which corresponds to the $n$-point correlation function of a Gaussian distribution. This means that $\rho (\zeta) $ is given by a leading Gaussian distribution with a non-Gaussian correction proportional to $\Lambda^4$. Thus, to find $\rho (\zeta)$ we may try the following ansatz
\be
\rho (\zeta) = \rho_G (\zeta) + \Delta \rho (\zeta) , 
\ee
where 
\be
\rho_G (\zeta) = \frac{e^{- \frac{\zeta^2}{2 \sigma_{\zeta}^2}}}{\sqrt{2\pi} \sigma_{\zeta}} ,
\ee
is the Gaussian part giving rise to the subset of $n$-point functions given in Eq.~(\ref{Gaussian-n-points}). The piece $\Delta \rho (\zeta) $ corresponds to the correction resulting from the nonlinear interactions proportional to $\Lambda^4$ (or equivalently, $A^2$). 

In what follows, we determine the form of  $\Delta \rho (\zeta) $ due to $\langle \zeta^n_L \rangle_c$ shown in Eq.~(\ref{n-point-final-res-coord-sol1}). The procedure crucially depends on knowing how $\langle \zeta^n_L \rangle_c$ depends on $n$, which requires us to deal with $I_n(\xi)$ of Eq.~(\ref{full-I_n}). To proceed, we find it useful to first show how to deduce $\Delta \rho (\zeta) $ in the case where $I_n(\xi)$ is given by its asymptotic form $I_n^0(\xi)$. This will then allow us to deal easily with the more general situation in which $I_n(\xi)$ is given by the full expression given in~(\ref{full-I_n}).

\subsection{Asymptotic reconstruction} \label{subsec:PDF-derivation-from-n-points}

If we take Eq.~(\ref{n-point-final-res-coord-sol1}) with $I_n (\xi)$ replaced by its asymptotic form $I_n^0(\xi)$ given in Eq.(\ref{asympt-I_n}), then $g_n = n$ when $n$ is even, and in this case we simply have
\bea
&& \langle \zeta^n_L \rangle_c =  (-1)^{n/2} n A^2 e^{- \frac{\sigma_\zeta^2}{2 f_\zeta^2}} \left[ \frac{\sigma_\zeta^2}{f_\zeta}   \right]^n  .  \label{zeta-n-2}
\eea
(and zero if $n$ is odd). This equation may be used to derive the PDF $\rho(\zeta)$ determining the probability of measuring a certain value of the curvature perturbation at an arbitrary position. To find $\Delta \rho (\zeta)$ we may try the following ansatz
\bea
\Delta \rho (\zeta) &=& \frac{e^{- \frac{\zeta^2}{2 \sigma_{\zeta}^2}}}{\sqrt{2\pi} \sigma_{\zeta}}  \Bigg[ \sum_{m=0} B_m \zeta^{2m} \cos\left( \frac{\zeta}{f_\zeta} \right)  \nn \\
&& \qquad + \sum_{m=0} C_m \zeta^{2 m +1}  \sin\left( \frac{\zeta}{f_\zeta}  \right) \Bigg]  , \label{PDF-ansatz}
\eea
where we have used the fact that $\rho(\zeta)$ must be even under the change $\zeta \to - \zeta$, for only the even moments are nonvanishing. It is direct to find that $B_0$ and $C_0$ are the only nonvanishing coefficients and, therefore, that the full PDF is given by~\cite{footnote-PDF-comp}
\be
\rho (\zeta) = \frac{e^{- \frac{\zeta^2}{2 \sigma_{\zeta}^2}}}{\sqrt{2\pi} \sigma_{\zeta}}  \left[ 1 + A^2   \frac{  \sigma_{\zeta}^2 }{ f_\zeta^2 }   \cos\left( \frac{\zeta}{f_\zeta} \right) - A^2  \frac{  \zeta }{ f_\zeta }   \sin\left( \frac{\zeta}{f_\zeta}  \right)  \right] . \label{PDF-1}
\ee
This PDF satisfies Eq.~(\ref{full-n-point}), and given that it corresponds to a small, absolutely continuous deformation of a Gaussian distribution, it is unique [that is, it is the only possible reconstruction from the moments of Eq.~(\ref{zeta-n-2})].

\begin{figure}[t!]
\includegraphics[scale=0.43]{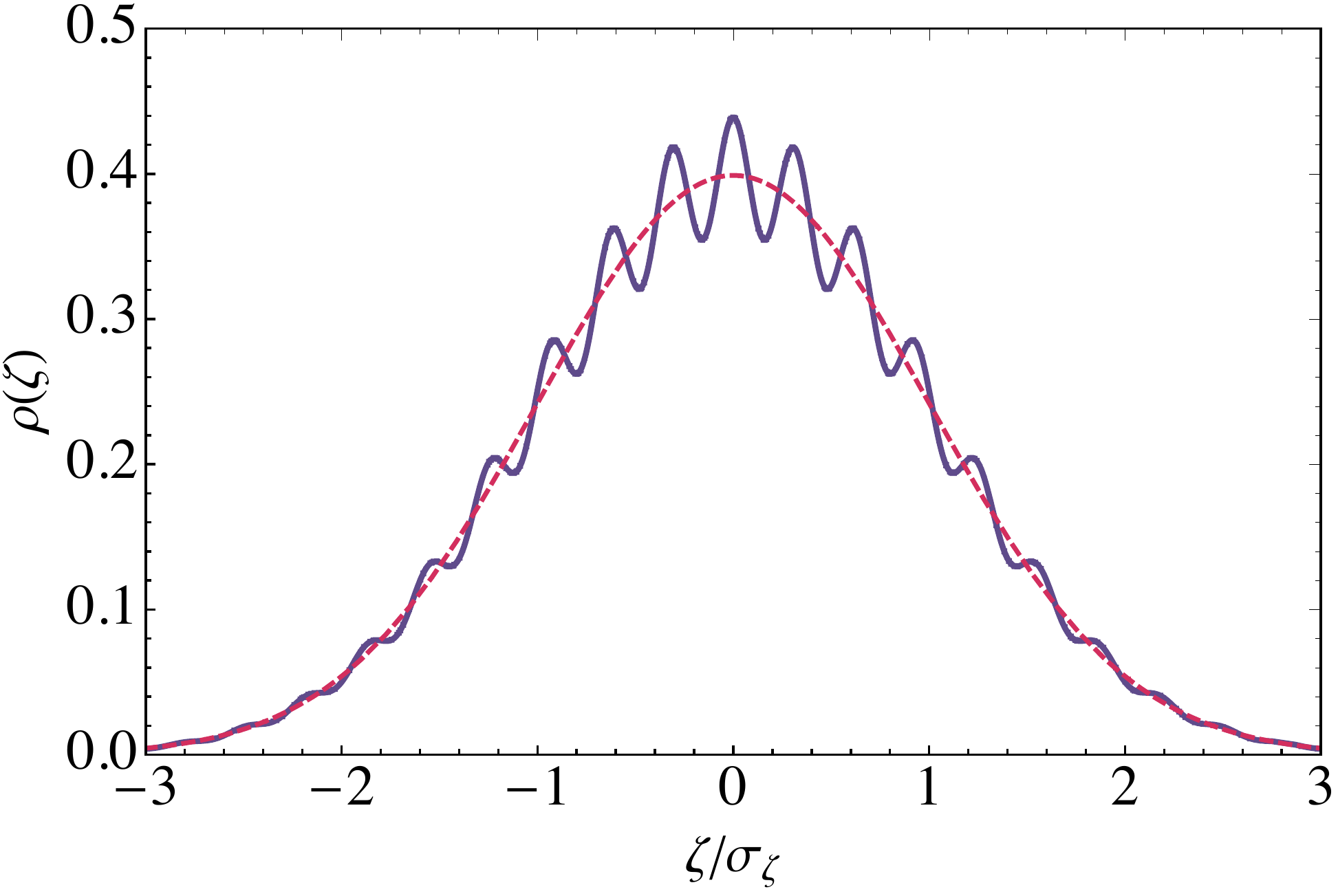}
\caption{An example of the PDF of Eq.~(\ref{PDF-1}) for the choice of parameters $f_\zeta / \sigma_\zeta = 5\times10^{-2}$ and $A^2 = 2.5 \times 10^{-4}$ (solid curve). For comparison, we have plotted a Gaussian distribution of variance $\sigma_\zeta$ (dashed curve).}
\label{fig:FIG-PDF}
\end{figure}

The probability distribution function~(\ref{PDF-1}) is valid in the formal limit $\xi \to \infty$. If $\sigma_L \gg H$ we could trust this result for the case $f \lesssim \sigma_L$, in which case the PDF shows nontrivial structures in the form of superimposed oscillations. However, given that $\sigma_L \sim H$ (because $\ln \xi \leq 60$), we cannot trust the regime  $f \lesssim \sigma_L$ (see Appendix~\ref{app_about_I_n}), and we are forced to consider the more general case in which $I_n$ is given by its full form shown in~(\ref{full-I_n}). In spite of this limitation, Eq.~(\ref{PDF-1}) constitutes one of our main results.  It gives a simple non-Gaussian probability distribution function for $\zeta$ in terms of various parameters related to the landscape shape. It may be verified that the PDF is already normalized as $\int \rho(\zeta) d\zeta = 1$. This probability distribution function is plotted in Fig.~\ref{fig:FIG-PDF} for specific values of $f_\zeta/ \sigma_\zeta$ and $A^2$. Notice that the second term inside the squared parenthesis in Eq.~(\ref{PDF-1}) accounts for the increase in probability of finding values of $\zeta$ that are sourced by those values of $\psi$ which minimize the cosine potential of Eq.~(\ref{potential-psi}). On the other hand, the third term, linear in $\zeta$ may be interpreted as a contribution accounting for the diffusion of $\zeta$ (one could in fact absorb the third term into the second term by slightly shifting $f_{\zeta}$).

\subsection{Full reconstruction} \label{subsec:PDF-derivation-from-n-points-full}

We now consider the task of deriving the full PDF, valid for any value of $\ln \xi > 0$. To proceed, it is helpful to realize that the most important aspect about the reconstruction performed in the previous section was the dependence of $\langle \zeta^n_L \rangle_c$ on $n$ as shown in Eq.~(\ref{zeta-n-2}). In the general case, if we consider the $x$ integral of Eq.~(\ref{full-I_n}) explicitly, we see that the dependence of $\langle \zeta^n_L \rangle_c$ on $n$ is exactly the same, except that this time it happens for each value of $x$. Then, a simple comparison with~\eqref{n-point-final-res-coord-sol1} shows that this time the reconstruction amounts to identifying an $x$ dependent decay constant 
\be
f_{\zeta}(x)\equiv f_{\zeta} \dfrac{\ln \xi }{F(\xi, x)} \geq f_{\zeta} , \label{f-x-def}
\ee
that satisfies $f_{\zeta}(0) = f_{\zeta}$. Hence, keeping track of all the numerical factors, we find
\bea
\rho (\zeta) = \frac{e^{- \frac{\zeta^2}{2 \sigma_{\zeta}^2}}}{\sqrt{2\pi} \sigma_{\zeta}}  \left[ 1 + A^2 \int_0^{\infty} \!\!\frac{dx}{x} \, \mathcal{K}_\xi(x)  \right. \qquad \qquad \nn \\  \left. \left(  \frac{  \sigma_{\zeta}^2 }{ f_\zeta(x)^2 }   \cos\left( \frac{\zeta}{f_\zeta(x)} \right) -  \frac{  \zeta }{ f_\zeta(x) }   \sin\left( \frac{\zeta}{f_\zeta(x)}  \right) \right)  \right] , \qquad  \label{PDF-full}
\eea
where the kernel $\mathcal{K}_\xi(x)$ is given by
\be
\mathcal{K}_\xi(x) \equiv \frac{ 2 G(\xi , x) \ln \xi }{\pi F(\xi, x)} \exp \left( -\frac{\sigma_{\zeta}^2 \left(f_{\zeta}^2(x) - f_{\zeta}^2\right)}{2 f_{\zeta}^2 f_{\zeta}^2(x)} \right). \label{def-kernel}
\ee
The result shown in Eq.~(\ref{PDF-full}) is our main result. It gives us the PDF for any value of the ratio $\xi =  k_{L}/k_{\rm IR} $. In particular, we can trust this result well inside the regime $f_\zeta < \sigma_\zeta$ for values $\ln \xi \simeq 8$, which corresponds to the range of scales available to CMB observations [as opposed to the case of the PDF of Eq.~(\ref{PDF-1})]. 

\begin{figure}[t!]
\includegraphics[scale=0.43]{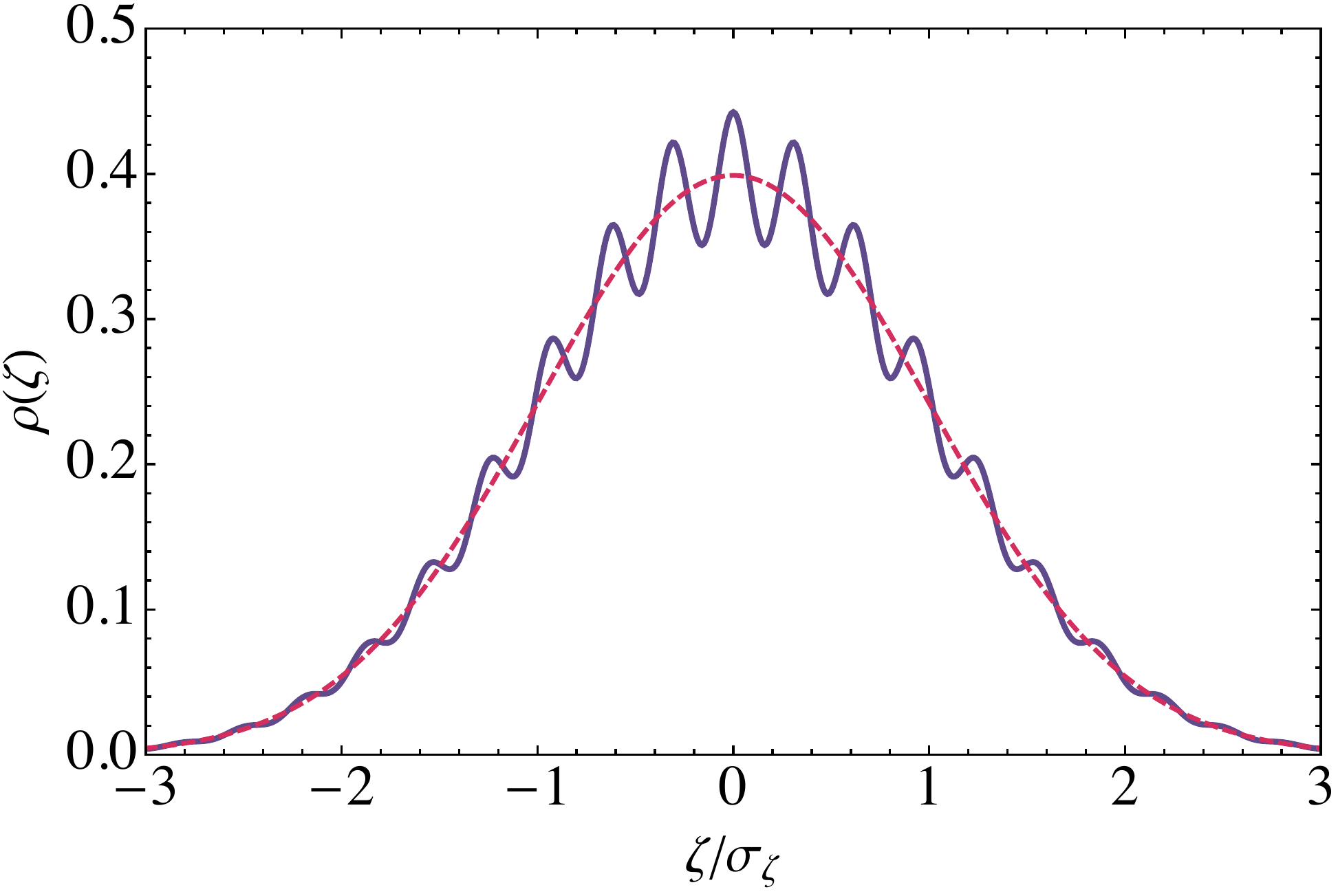}
\caption{An example of the PDF of Eq.~(\ref{PDF-full}) for the choice of parameters $f_\zeta / \sigma_\zeta = 5 \times10^{-2}$, $A^2 =  2.5 \times 10^{-2}$ and $\ln \xi = 8$ (solid curve). For comparison, we have plotted a Gaussian distribution of variance $\sigma_\zeta$ (dashed curve). Notice that $A^2$ is 100 times larger than the value used to plot Fig.~\ref{fig:FIG-PDF}.}
\label{fig:FIG-PDF-1}
\end{figure}

\begin{figure}[t!]
\includegraphics[scale=0.43]{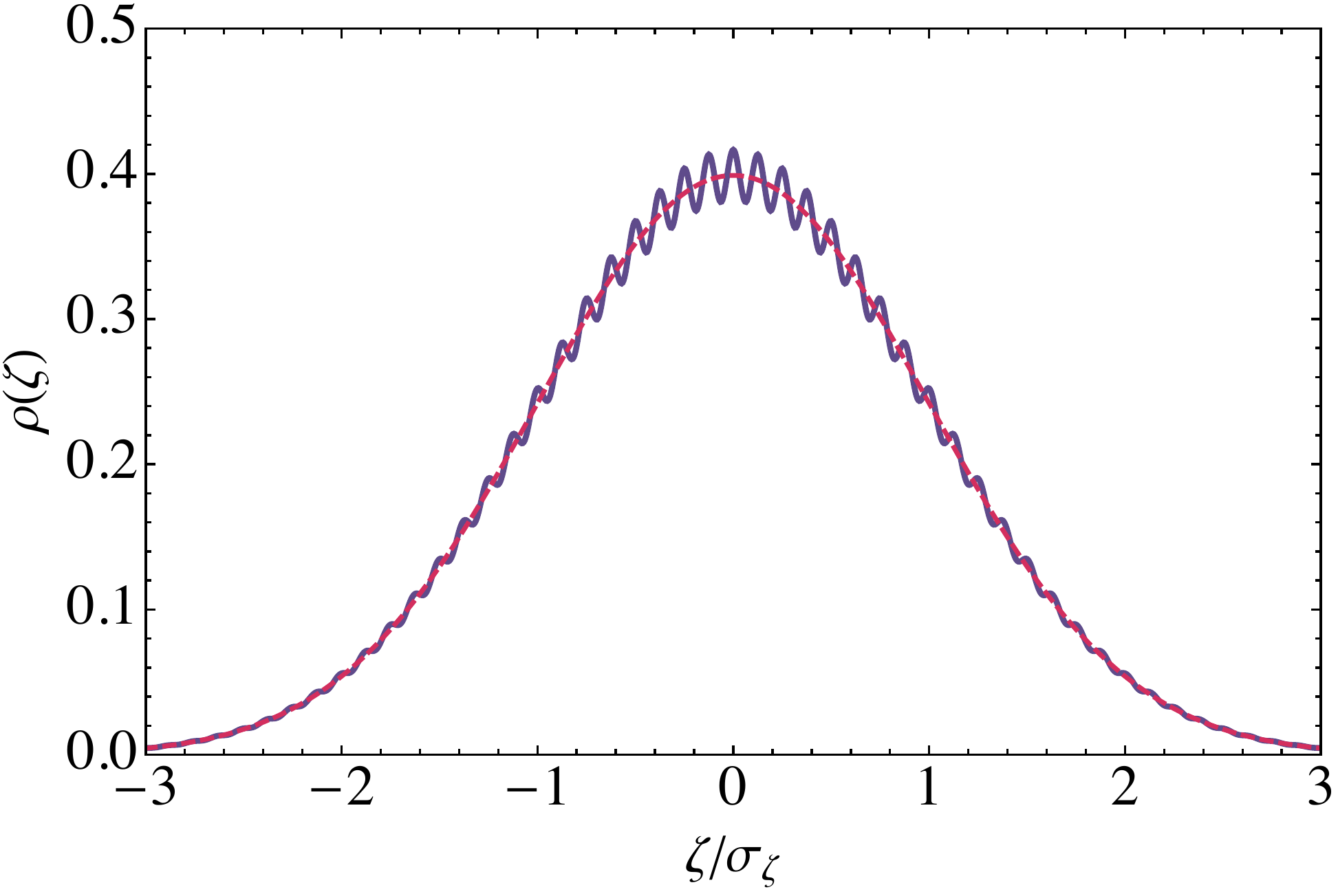}
\caption{An example of the PDF of Eq.~(\ref{PDF-full}) for the choice of parameters $f_\zeta / \sigma_\zeta = 2 \times10^{-2}$, $A^2 =  2.5 \times 10^{-2}$ and $\ln \xi = 8$ (solid curve). For comparison, we have plotted a Gaussian distribution of variance $\sigma_\zeta$ (dashed curve). Here $A^2$ has the same value as that of Fig.~\ref{fig:FIG-PDF-1}, but $f_\zeta / \sigma_\zeta$ is smaller.}
\label{fig:FIG-PDF-2}
\end{figure}

An outstanding property of~(\ref{PDF-full}) is that it preserves the oscillatory structure of the potential in a strikingly similar manner as~\eqref{PDF-1}. The main difference, is that now there is a filtering function that accounts for the effects that arise when one considers only the bounded region of $k$ space which we are able to probe. The consequences of this filtering can be appreciated by looking at Fig.~\ref{fig:FIG-PDF-1} (plotted for $\ln \xi = 8$). There we see by comparison to Fig.~\ref{fig:FIG-PDF} that the amplitude of the oscillations in the full PDF is suppressed, since the value chosen for $A^2$ in this last plot is $100$ larger than the previous one. Moreover, as illustrated in Fig.~\ref{fig:FIG-PDF-2} (also plotted for $\ln \xi = 8$), decreasing the value of $f_{\zeta}$ from this point does not enhance the amplitude as we might have thought while looking at Eq.~(\ref{PDF-1}) but the opposite: the amplitude actually gets smaller, mostly because of the exponential factor in the kernel $\mathcal{K}_\xi(x)$.

Note that in the formal limit $\ln  \xi \to \infty$ the probability distribution function~(\ref{PDF-full}) becomes independent of $\xi$, and we recover~\eqref{PDF-1}. This corresponds to the ideal situation whereby the entire range of momenta is available to observers. In that case, one can directly infer the parameters $\Lambda^4_{\rm ren}$ and $f$ of the landscape potential by performing statistics with observations of primordial curvature perturbations. Otherwise, as long as observers can only have access to a limited amount of modes, as parametrized by $\ln \xi$, the filtering function appearing in the kernel $\mathcal K_\xi(x)$ will wash out the structure of the potential. This is simply because $\ln \xi$ restricts the number of modes in momentum space that can add up to increase the effect of nonlinearities due to $\Delta V$ on the PDF in coordinate space: momentum conservation through the vertex implies that  while certain modes are probing large (small) momenta limited by $q_L$ ($q_{\rm IR}$), other momenta will probe more restricted regions in momentum space. As a result, to extract information about $\Delta V$ one has to take into account the role of $\ln \xi$. We will come back to this issue in Ref.~\cite{Chen:2018brw}.

\section{Discussion and conclusions} \label{sec:conclusions}

We have examined a regime of multifield inflation where the shape of the landscape potential in the isocurvature direction can be probed using non-Gaussianity of primordial density perturbations. At the level of $n$-point correlation functions (or polyspectra), these non-Gaussianities take the local form, as in all multifield models \cite{Enqvist:2004bk, Lyth:2005fi, Seery:2005gb, Rigopoulos:2005ae, Alabidi:2005qi, Battefeld:2007en, Byrnes:2008wi, Byrnes:2009qy, Battefeld:2009ym, Elliston:2011et, Mulryne:2011ni, McAllister:2012am, Byrnes:2012sc, Bjorkmo:2017nzd, Langlois:1999dw, Gordon:2000hv, GrootNibbelink:2000vx, GrootNibbelink:2001qt, Amendola:2001ni, Bartolo:2001rt, Wands:2002bn, Tsujikawa:2002qx, Byrnes:2006fr, Choi:2007su, Cremonini:2010sv, Cremonini:2010ua, Achucarro:2016fby} and quasi-single-field models \cite{Chen:2009we,Chen:2009zp, Baumann:2011nk, Chen:2012ge, Sefusatti:2012ye, Norena:2012yi, Noumi:2012vr, Gong:2013sma, Emami:2013lma, Kehagias:2015jha, Arkani-Hamed:2015bza,  Dimastrogiovanni:2015pla,Chen:2015lza, Chen:2016cbe,Lee:2016vti,Chen:2016qce, Meerburg:2016zdz,Chen:2016uwp,Chen:2016hrz, An:2017hlx,Iyer:2017qzw,An:2017rwo, Kumar:2017ecc, Franciolini:2017ktv,Tong:2018tqf, MoradinezhadDizgah:2018ssw, Saito:2018xge,Franciolini:2018eno,Chen:2018sce} with sufficiently light isocurvatons. However, a novel point of this paper is that the information about the shape of the potential is not manifest in the individual $n$-point correlation functions, but rather in the resummed probability distribution function given in Eq.~(\ref{PDF-full}). This shows that local non-Gaussianity may have a rich structure inherited by the self-interactions of isocurvature fields, together with a derivative coupling common to multifield models.

Although the mechanism of statistical transfer examined in this article is based on the derivative coupling $\mathcal L_{\rm int} \propto \dot \zeta \psi$, our results are likely to be more general. We therefore expect that other classes of interactions between the curvature perturbation and other scalar fields scanning the landscape lead to similar conclusions.

Also, as mentioned in the Introduction, the particular form of the potential $\Delta V(\psi)$ should not be so crucial. While it is true that the cosine function used in this work comes with the right properties making the loop resummations possible, more general potentials are in fact not intractable. In a companion Letter~\cite{Chen:2018brw}, we will extend Eq.~(\ref{PDF-full}) to the case of more general potentials $\Delta V(\psi)$ and analyze the possibility of reconstructing such potentials out of observable quantities.

In what follows, we discuss various relevant aspects related to our main result.

\subsection{Aspects of nonperturbativity}

The probability distribution function of Eq.~(\ref{PDF-full}) was the result of two resummations. The first resummation corresponded to the loop expansion coming from the infinite number of vertices following from the Taylor expansion of Eq.~(\ref{cosine-expansion}). This makes our PDF nonperturbative in terms of the parameter $H/f$ (or, equivalently, $\sigma_L/f$ and thus $\sigma_{\zeta}/f_{\zeta}$). The second resummation corresponded to the derivation of the PDF itself, from the infinite set of $n$-point functions given by (\ref{zeta-n-1}). This second resummation gave us back the appearance of the cosine function, weighted by the kernel $\mathcal{K}_\xi(x)$ of Eq.~(\ref{def-kernel}) and with a periodicity determined by $f_\zeta (x)$ as in Eq.~(\ref{f-x-def}). These two functions are such that the non-Gaussian correction to the PDF preserves the structure of the landscape potential $\Delta V$.

These two steps (the resummations) work in tandem and inform us about some crucial aspects of our result. While each $n$-point function depends nonperturbatively on $\sigma_{\zeta}/f_\zeta$, by themselves they cannot give an account of the oscillatory structure of the PDF. In particular, the lowest non-Gaussian correlation function (the four-point function) can hardly inform us about the nature of the class of non-Gaussianity from where it is derived. This is the most salient point of this work: in order to correctly describe the nonlinear effects of the isocurvature field on the curvature perturbation $\zeta$, we cannot just limit ourselves to the lowest non-Gaussian $n$-point correlation function.

As discussed in Sec.~\ref{sec:tom-NG}, observables depend on the renormalized parameter $\Lambda^4_{\rm ren} =e^{- \sigma_{S}^2 / 2 f^2} \Lambda^4$, which is a result of the loop resummation. Given that this resummation will reappear at any order of the expansion with respect to $\Lambda^4$, we should take the perturbative expansion of the hight of the potential as being controlled by the renormalized parameter $\Lambda^4_{\rm ren} / H^4$ instead of $\Lambda^4 / H^4$. In other words, higher order corrections to the PDF of Eq.~(\ref{PDF-full}) are expected to be of order $A^2$.

It is important to stress that, while our result is perturbative in $\Lambda^4_{\rm ren} / H^4$ and $\lambda$, the main features of the PDF (\ref{PDF-full}) should persist to other regimes. For example, we have stayed in the region $\lambda \ll 1$ in order to be able to treat the mixing term (\ref{H-mixing}) as part of the interaction Hamiltonian. This regime has the limitation that, in order for the effects coming from $\lambda$ to be relevant, one needs a large number of $e$-folds $\Delta N$ after horizon crossing. On the other hand, we expect that the qualitative characteristics found in the PDF of Eq.~(\ref{PDF-full}) springing from $\lambda$ will be enhanced in the case $\lambda \gtrsim 1$.

\subsection{Relation to previous works}

Our analysis has some similarities (but also important differences) with previous works studying the implications of isocurvature fields on the production of primordial non-Gaussianity. In quasi-single-field inflation models, the isocurvature field is assumed to be massive and to have some interactions, such as the cubic self-interaction as in Eq.~(\ref{hierarchical-exp-V})
\be
\Delta V(\psi) = \frac{1}{2} \mu^2 \psi^2 + \frac{1}{6} g \psi^3 .  \label{QSF-pot}
\ee
In these models, the mass of $\psi$ in principle can be a free parameter and plays an important role: it controls the extent to which the fluctuations of $\psi$ can survive at superhorizon scales and interact with $\zeta$. This is because the amplitude of $\psi$ decays after horizon crossing as
\begin{equation}
\psi \sim e^{- \frac{3}{2} (1 - R) \Delta N},
\end{equation}
where $R$ is the real part of $ \sqrt{1 - 4 \mu^2 /9 H^2}$.

Although many results apply for general $\mu$, the most interesting cases in the quasi-single-field literature \cite{Chen:2009we,Chen:2009zp,Baumann:2011nk,Chen:2012ge, Sefusatti:2012ye, Norena:2012yi, Noumi:2012vr, Gong:2013sma, Emami:2013lma, Kehagias:2015jha, Arkani-Hamed:2015bza,  Dimastrogiovanni:2015pla,Chen:2015lza, Chen:2016cbe,Lee:2016vti,Chen:2016qce, Meerburg:2016zdz,Chen:2016uwp,Chen:2016hrz,
An:2017hlx,Iyer:2017qzw,An:2017rwo, Kumar:2017ecc, Franciolini:2017ktv,Tong:2018tqf, MoradinezhadDizgah:2018ssw, Saito:2018xge,Franciolini:2018eno,Chen:2018sce} are those with $\mu \sim H$.
Since the potential \eqref{QSF-pot} is assumed to hold within a field range much larger than the amplitude of $\psi$, ${\cal O}(H)$,
the isocurvaton field $\psi$ is confined within this potential and does not fluctuate outside to explore the fuller structure of the landscape. In fact, the PDF of the density perturbation of the quasi-single-field inflation model may be worked out in a similar fashion and it should encode the shape of the potential, although it has much less rich structure.
The main predictions for non-Gaussianities coming from~(\ref{QSF-pot}) are some nontrivial polyspectra, such as bispectra and trispectra.

In the case studied in this article, the field $\psi$ does not decay. Note that, classically, $\psi$ has a mass coming from the cosine potential of Eq.~(\ref{potential-psi}),
\be
\mu^2 = \frac{\Lambda^4}{f^2} ,
\ee
and this quantity may be even larger than $H^2$. But given that the potential barriers are small ($\Delta V^{1/4} \sim \Lambda \ll H$ at tree level) the $\psi$ fluctuations will still be able to traverse vigorously the barriers of the potential and explore the potential landscape.  In other words, the $\psi$ field is effectively massless at the leading order.
In this case, the classical mass term is only part of the rich structure in the small perturbation $\Delta V$ and appears as the first term in the series expansion of $\Delta V$. Therefore, this series expansion needs to be resummed in the final result.
In the case of the cosine potential studied here, these aspects are summarized in the fact that all the vertices depend on just two parameters ($\Lambda^4$ and $1/f$), and so every vertex contributes decisively in the computation of the $n$-point correlation functions.

\subsection{After inflation}

So far we have established how a scalar field $\psi$ with a non-Gaussian distribution function can transfer its statistics to the curvature perturbation $\zeta$ during inflation. The mechanism by which the statistics is transferred is cumulative: $\psi$ transfers its statistics (both Gaussian and non-Gaussian) to $\zeta$ as long as $\lambda \neq 0$, and the non-Gaussianity of $\zeta$ becomes more accentuated as time passes. After a long enough period, the statistics of $\zeta$ (which in the absence of the $\lambda$ coupling would be nearly Gaussian) becomes completely dominated by that of the curvature perturbation $\psi$.

Three main things could happen after such a period that bring this mechanism to an end:
(1) As mentioned, if $\psi$ is not exactly massless, after some $e$-folds at superhorizon it will naturally decay, as in quasi-single-field inflation models.
(2) The potential $\Delta V(\psi)$ changes drastically. In more realistic scenarios we would expect that the potential $\Delta V(\psi)$ depends explicitly on time due to its dependence on the background. Before the end of inflation $\Delta V(\psi)$ could introduce a new scale that makes $\psi$ a very massive field within the relevant amplitude range $\sigma_L \simeq H$. In that case, the amplitude of $\psi$ would quickly decay (due to the kinematics of a massive field in an expanding Universe) and $\psi$ would not be able to source $\zeta$ any more. (3) The third possibility is that $\lambda$ effectively vanishes before the end of inflation [before even $\Delta V(\psi)$ changes]. Here, even though the amplitude of $\psi$ has not decayed, the sourcing offered by $\psi$ ends.

In all of the previous cases, the non-Gaussian statistics of $\zeta$ persists, simply because on superhorizon scales $\zeta$ remains constant (after $\psi$ has done its job of sourcing its statistics). In other words, the statistics transferred by $\psi$ while $\lambda \neq 0$ and $\psi \neq 0$ serves as the initial condition for a posterior phase where $\lambda = 0$ and/or $\psi = 0$. Thus, because the statistical transfer is cumulative, the new phase with $\lambda = 0$ and/or $\psi = 0$ would not imply that the non-Gaussian statistics of $\zeta$ is erased. All the contrary, if $\lambda =0$ or $\psi$ becomes massive, then $\zeta$ would kinematically decouple from $\psi$ and would continue to evolve independently, with a frozen amplitude, preserving its non-Gaussian statistics. Of course, the statistics of $\zeta$ would then survive reheating until horizon entry, fixing the initial conditions for perturbations in the hot Big-Bang era.

\subsection{Current constraints on the PDF}

We can constrain the level of non-Gaussianity in the probability distribution function~(\ref{PDF-full}) by looking into current bounds on the trispectrum~\cite{Okamoto:2002ik, Boubekeur:2005fj, Alabidi:2005qi, Kogo:2006kh, Seery:2006vu, Seery:2006js} set by Planck, as this model has an identically vanishing bispectrum [see Eq.~\eqref{n-point-final-res-zeta}], and consequently, we cannot use it to constrain the resulting PDF.

\begin{figure}[t!]
\includegraphics[scale=0.43]{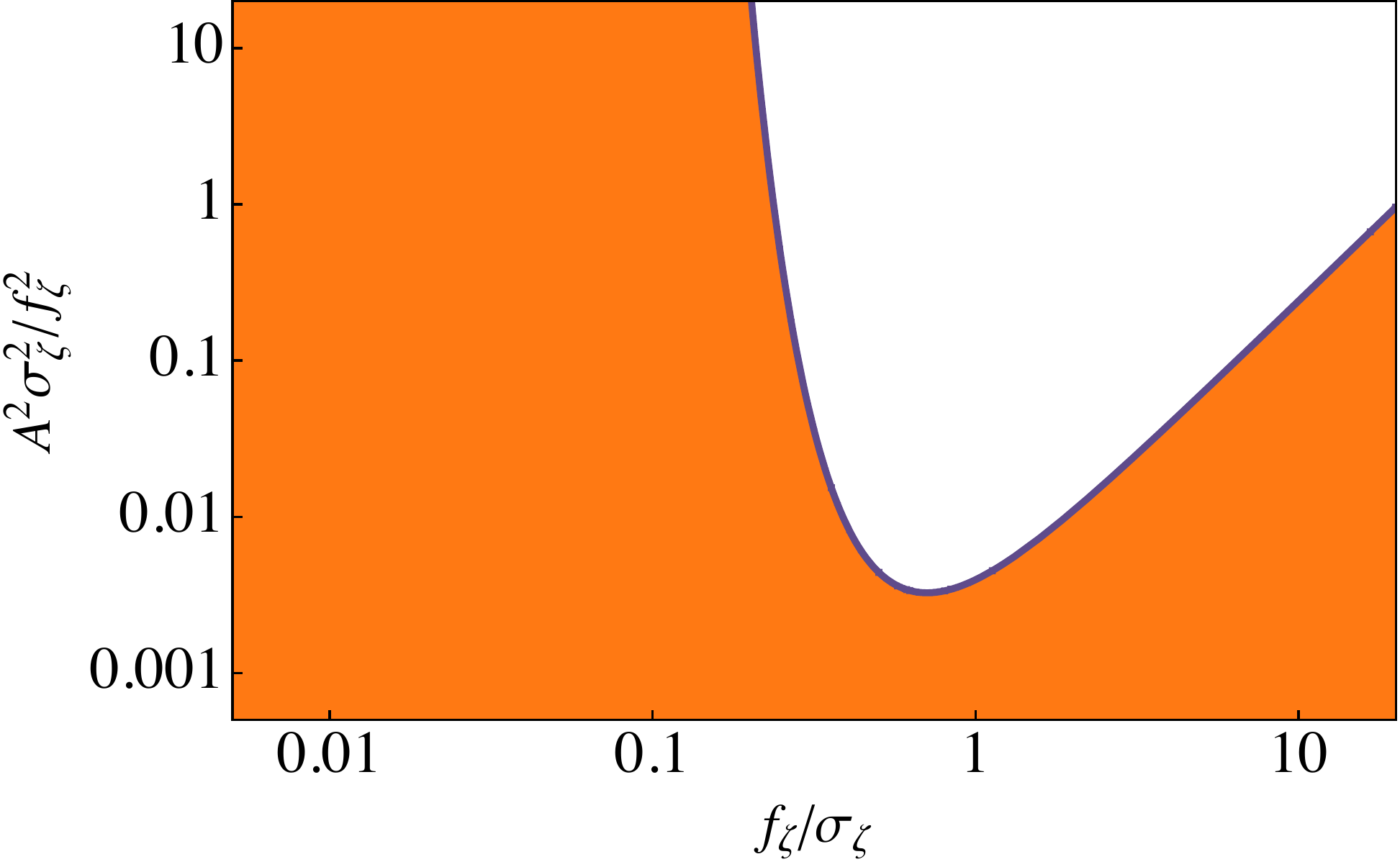}
\caption{The allowed values for the parameters $A^2$ and $f_\zeta / \sigma_\zeta$ deduced from current constraints by Planck on the trispectrum (at 95\% C.L.). The allowed region (in orange) is located bellow the solid curve. Notice that the combination $A^2 \sigma_\zeta^2 / f_\zeta^2$ corresponds to the coefficient in front of the cosine function in Eq.~(\ref{PDF-full}) at $x=0$.}
\label{fig:FIG-TRI}
\end{figure}

Specifically, Planck is able to constrain the parameter $g_{\rm NL}^{\rm local}$ that appears in the following relation involving the four-point function for $\zeta$, and its power spectrum:
\bea
\langle \zeta_{\k_1}\zeta_{\k_2}\zeta_{\k_3}\zeta_{\k_4} \rangle \equiv (2\pi)^3\delta^{(3)}\left(\sum \k_{i} \right)\frac{54}{25}g_{\rm NL}^{\rm local} \nn \\
 \times \left[P_{\zeta}(k_1)P_{\zeta}(k_2)P_{\zeta}(k_3)+3 \, \text{perm.} \right].
\eea
This expression may be compared with our general expression~(\ref{n-point-final-res-zeta}) for the specific case $n=4$, which is given by
\bea
&& \langle \zeta_{\k_1}\zeta_{\k_2}\zeta_{\k_3}\zeta_{\k_4} \rangle =   (2 \pi)^3 \delta^{(3)} \Big(\sum_j \k_j \Big) \frac{\Lambda^4}{3H^4} \nn \\
&& \quad e^{- \frac{\sigma_0^2}{2 f^2}} \left( \frac{\lambda H^2 \Delta N}{2 f \sqrt{2 \epsilon}} \right)^4 \left[ \frac{1}{k_1^3 k_2^3 k_3^3}  +3 \, \text{perm.}  \right]\Delta N . \,\,\,\,\,\,\,\,\,\,
\eea
To compare both expressions, it is necessary to recall, from the discussion around Eq.~(\ref{Pzeta-Ppsi}), that the power spectrum for $\zeta$ is given by
\be
P_\zeta (k) =  \frac{\lambda^2 H^2 \Delta N^2}{4  \epsilon k^3 } . \label{power-spectrum-zeta}
\ee
Then, it follows that $g_{\rm NL}^{\rm local}$ is given by
\bea
g_{\rm NL}= \frac{25}{54} \frac{A^2 (2\epsilon)}{\lambda^2\Delta N^2}  \frac{\sigma_{L}^2}{f^4}  e^{- \frac{\sigma_L^2}{2 f^2}}.
\eea
We can turn this expression into a more useful result by recalling, from Eqs.~(\ref{sigma-zeta}) and (\ref{f-sigma-zeta}), that $\sigma_\zeta^2  = \sigma_L^2 \lambda^2 \Delta N^2 / 2 \epsilon$ and $f_{\zeta} / \sigma_{\zeta} = f / \sigma_L$. We obtain
\bea
g_{\rm NL}= \frac{25}{54} A^2 \frac{ \sigma_{\zeta}^2}{f_{\zeta}^4}  e^{- \frac{\sigma_\zeta^2}{2 f_{\zeta}^2}}.
\eea
With the help of Eq.~(\ref{sigma-cut-off}) we see that $\sigma_{\zeta}^2$ is related to the power spectrum (\ref{power-spectrum-zeta}) through
\be
\sigma_{\zeta}^2 =   \frac{P_\zeta (k) k^3}{2 \pi^2} \ln \xi ,
\ee
(recall that $\xi = q_L / q_{\rm IR}$). Planck observations~\cite{Ade:2015lrj} currently constrain the amplitude of the power spectrum as $P_\zeta (k) k^3 / 2 \pi^2 = (2.196 \pm 0.158) \times 10^{-9}$. Then, by setting $\ln \xi = 8$, the range of scales corresponding to the CMB, we may write $\sigma_{\zeta}^2 \simeq 1.3 \times 10^{-7}$. Then $g_{\rm NL}$ becomes
\bea
g_{\rm NL} \simeq 3.5 \times 10^6 \, A^2 \frac{ \sigma_{\zeta}^4}{f_{\zeta}^4}  e^{- \frac{\sigma_\zeta^2}{2 f_{\zeta}^2}}.
\eea
Furthermore, current constraints on the primordial trispectrum by Planck~\cite{Ade:2015ava} imply $g_{\rm NL} = ( - 9.0 \pm 15.4) \times 10^4$ at $95$\% C.L. It then follows that $A^2$ and the ratio $ f_{\zeta}^2 / \sigma_\zeta^2$ must satisfy the following restriction:
\be
A^2 \frac{ \sigma_{\zeta}^4}{f_{\zeta}^4}  e^{- \frac{\sigma_\zeta^2}{2 f_{\zeta}^2}} < 2.1 \times 10^{-3} .
\ee
Figure~\ref{fig:FIG-TRI} shows the allowed values for the parameter space spanned by $A^2$ and $f_\zeta / \sigma_\zeta$. It may be seen that $A^2$ becomes less constrained for both, large and small values of $f_\zeta / \sigma_\zeta$. It is interesting to note that the values used to plot both Fig.~\ref{fig:FIG-PDF-1} ($f_\zeta / \sigma_\zeta = 5 \times 10^{-2}$ and $A^2 \sigma_\zeta^2 / f_\zeta^2 = 10$) and Fig.~\ref{fig:FIG-PDF-2} ($f_\zeta / \sigma_\zeta = 2 \times 10^{-2}$ and $A^2 \sigma_\zeta^2 / f_\zeta^2 = 62.5$) are well within the allowed region. However, it is hard to conceive that peaks in the non-Gaussian PDF larger than those shown in Fig.~\ref{fig:FIG-PDF-1} are not excluded. This suggests that even strong constraints on the four-point function would not compete with other methods aiming to constrain the shape of the probability distribution function. 

\subsection{Concluding remarks}

To conclude, we have studied the generation of a novel class of non-Gaussianity, in multifield inflation with the help of a well-motivated model involving an axionlike field coupled to the comoving curvature perturbation. We showed that the isocurvature fluctuations can traverse the periodic potential by fluctuating across the barriers at the time of horizon crossing. As a result, the isocurvature field acquires rich statistics that reflect the landscape structure. These statistics may be transferred to the adiabatic mode via a derivative coupling which is the lowest possible coupling that arises from an effective field theory point of view (a concrete UV completion of this effective field theory can be found in Appendix~\ref{sec:concrete-example}).

The outcome of nonperturbative resummations is a probability distribution function for the curvature perturbation which consists of a Gaussian part corrected by a term involving information about the landscape potential. In such a case, the traditional perturbative $f_{\rm NL}$ parametrization cannot probe the full non-Gaussian structure and new estimators probing the whole PDF of temperature anisotropies should be developed.

Our study gives an example where a stringy landscape may have calculable effects on the CMB and large-scale structure that have not yet been thoroughly searched for in the current data. Last but not least, our result could have important implications for understanding the generation of primordial black holes~\cite{Franciolini:2018vbk}.

\begin{acknowledgments}

We are grateful to Ana Ach\'ucarro, Rafael Bravo, Sebasti\'an C\'espedes, Anne C. Davis, Jorge Nore\~na, Dong-Gang Wang, Yvette Welling and Matias Zaldarriaga for useful discussions and comments. XC is supported in part by the NSF Grant No. PHY-1417421. GAP, WR and BSH acknowledge support from the Fondecyt Regular Project No. 1171811 (CONICYT). BSH is supported by a CONICYT Grant No. CONICYT-PFCHA/Mag\'{i}sterNacional/2018-22181513. SS is supported by the Fondecyt Postdoctorado Project No. 3160299 (CONICYT).

\end{acknowledgments}

\begin{appendix}

\section{A concrete example} \label{sec:concrete-example}

Here we describe a concrete example of an ultraviolet complete model from where the Lagrangian for the fluctuations $\zeta$ and $\psi$ shown in Eq.~(\ref{Lagrantian-full-v}) emerges. Our starting point is to consider the following general multifield action for the set of fields $\phi^a$, with $a=1,2$:
\be
S = \frac{1}{2} \int d^4 x R -  \int d^4 x \Big[ \frac{1}{2}  \gamma_{a b} \partial_{\mu} \phi^{a} \partial^{\mu} \phi^{b}  - V (\phi)  \Big] , \label{action-phi-a}
\ee
where $\gamma_{a b}$ is a sigma model metric describing the geometry of the target field space. Before we specify a particular model by choosing $\gamma_{a b}$ and $V (\phi)$, it is useful to recall the following property found in holographic models of inflation~\cite{Garriga:2014fda}: if the potential $V (\phi)$ can be written in terms of a function $W(\phi)$ as
\be
V (\phi) = 3 W^2 - 2 \gamma^{ab} W_a W_b , \label{V-W}
\ee
(with $W_a = \partial_a W$) then the system admits the following background solution:
\bea
H &=& W , \\
\dot \phi^a &=& - 2 \gamma^{a b} W_b .
\eea
We will exploit these relations in a moment. First, let us set the field content as $\phi^1 = \X$, $\phi^2 = \Y$, and write the sigma model metric as
\be
\gamma_{ab} = \left(\begin{array}{cc} e^{2 \Y / R_0} & 0 \\0 & 1\end{array}\right) . \label{metric-XY}
\ee
The nonvanishing Christoffel symbols are found to be $\Gamma^{\X}_{\Y\X} = \Gamma^{\X}_{\X\Y} = 1 / R_0$ and $\Gamma^\Y_{\X\X} = - e^{2 \Y / R_0} /R_0$. Then, it is direct to verify that the metric (\ref{metric-XY}) describes a hyperbolic target space with a Ricci scalar given by $\mathbb{R} = - 2 / R_0$. With this form of the metric, the background equations of motion are found to be given by
\bea
&& 3 H^2 = \frac{1}{2} e^{2 \Y / R_0} \dot \X^2 + \frac{1}{2} \dot \Y^2  + V  , \\
&& \ddot \X + 3 H \dot \X + \frac{2}{R_0} \dot \Y \dot \X +  e^{- 2 \Y / R_0  } \partial_\X V  = 0 , \\
&& \ddot \Y + 3 H \dot \Y - \frac{1}{R_0} e^{ 2 \Y / R_0  } \dot \X^2  + \partial_\Y V  = 0 .
\eea
Let us now specify the potential $V (\X,\Y )$ by splitting it as
\be
V(\X,\Y ) = V_0 (\X,\Y ) + \Delta V(\Y)  .
\ee
The term $V_0 (\X,\Y )$ corresponds to the main contribution driving inflation, whereas $\Delta V (\Y)$ will give rise to the self-interaction of the fluctuation $\psi$. We need to set $V_0 (\X,\Y )$ appropriately so that the action for the perturbations reduces to (\ref{Lagrantian-full-v}). We choose $V_0 (\X,\Y )$ by demanding that it has the structure shown in (\ref{V-W}), with a function $W = W(\X)$ (a function of $\X$ only). That is, we write
\be
V_0 (\X,\Y ) = 3 W^2 - 2 e^{-2 \Y / R_0} W_\X^2 .
\ee
Now, it is a simple matter to show that the system admits the following background solution:
\bea
&& H = W, \label{bg-1} \\
&& \dot \X = - 2 e^{- 2 \Y / R_0  }  W_\X , \label{bg-2} \\
&& \dot \Y = 0,  \label{bg-3}
\eea
as long as the constant value of $\Y$ coincides with a local minimum of $\Delta V (\Y)$, such that $\Delta V (\Y) = 0$, and
\be
 \frac{\partial}{\partial \Y }\Delta V (\Y) = 0 . \label{bg-4}
\ee
That is, the field $\X$ evolves and drives inflation while $\Y$ stabilizes to a local minimum of $\Delta V (\Y)$.

This background solution is characterized by an inflationary trajectory in field space with tangent and normal vectors $T^a$ and $N^a$ given by
\be
T^a = ( e^{- \Y / R_0} , 0 ) , \quad N^a = (0 , 1) .
\ee
Then, one may verify that the covariant time variation of $T^a$ is linked to $N^a$ through the following equation
\be
\frac{D}{dt} T^a = - \Omega N^a,
\ee
where the covariant time derivative $D/dt$ is defined to act on an arbitrary vector $V^a$ as $D V^a / dt = \dot V^a + \Gamma^{a}_{bc} V^b \dot \phi^c$.  The quantity $\Omega$ corresponds to the rate of turn (angular velocity) characterizing the bend of the inflationary trajectory. In the present case, it reads
\be
 \Omega =  - \frac{2}{R_0} e^{- \Y / R_0}  W_\X .
\ee
Thus, we see that in this model the background dynamics consists of an inflationary background with a nonvanishing rate of turn $\Omega$. To have the appropriate amount of inflation, one needs to choose $W(\X)$ appropriately, but the details of this choice are not relevant for this discussion.

One may now perturb the system as $\phi^a (\x , t) = \phi_0^a + \delta \phi^a  (\x , t)$. In other words, we may write
\bea
\X (\x,t) &=& \X_0(t) + \delta \X (\x,t) ,  \\
\Y (\x,t) &=& \Y_0 + \psi (\x,t) ,
\eea
where $\X_0(t)$ and $\Y_0$ are the background fields satisfying (\ref{bg-1})-(\ref{bg-4}).

To find the perturbed system we choose the comoving gauge, whereby the fluctuation along the inflationary trajectory is set to vanish $T_{a} \delta \phi^a = 0$. This is equivalent to $\delta \X (\x,t) = 0$. In this gauge, the relevant fluctuations are the field fluctuation normal to the trajectory $N_{a} \delta \phi^a = \psi$ and the comoving curvature perturbation $\zeta$, which is introduced through the following perturbed metric written with the help of the ADM decomposition:
\be
ds^2 = - N^2 dt^2 + a^2 (t) e^{2 \zeta} \delta_{ij} (dx^i + N^i dt) (dx^j + N^j dt) ,
\ee
where $N$ and $N^i$ are the usual lapse and shift functions. It is now straightforward to deduce the action for the fluctuations $\zeta$ and $\psi$. If one keeps $\zeta$ and $\psi$ quadratic everywhere in the action, except for $\Delta V$, where one keeps $\psi$ to all orders, then one obtains the Lagrangian of Eq.~(\ref{Lagrantian-full-v}). The parameter $\alpha$ of Eq.~(\ref{Lagrantian-full-v}) is related to $\Omega$ by
\be
\alpha =  \frac{\Omega}{\sqrt{2 \epsilon}} .
\ee
 As a final step, the potential (\ref{potential-psi}) is obtained for the particular choice
\be
\Delta V (\Y) = \Lambda^4 \left[ 1 - \cos \left( \frac{\Y}{f} \right) \right] . \label{pot-sin}
\ee
The reason why we have kept $\psi$ up to quadratic order everywhere in the action except for $\Delta V$ is that we are interested in treating the range of field fluctuations in $\Delta V$ nonperturbatively. In the case of (\ref{pot-sin}), this is equivalent to treating $1/f$ nonperturbatively.

\section{Details of computations} \label{sec:details}

Here we provide details of the intermediate steps of the computation discussed in Sec.~\ref{sec:npoints} leading to Eq.~(\ref{n-point-final-res-zeta}). Our starting point consists of plugging the field $u$, expanded in terms of the interaction picture field $u_I$ shown in Eq.~(\ref{u-u_I}), back into $\langle \hat u (\k_1, \tau)  \cdots \hat  u(\k_n , \tau) \rangle $. We obtain
\begin{widetext}
\bea
\langle u(\k_1,\tau)\cdot \cdot \cdot u(\k_n,\tau) \rangle_{\rm c} =
\sum_{j = 1}^n \sum_{p = 0}^{n-1} \sum_{I_p} i^{n+1} \bra{0} \left(\prod_{\substack{l = 1 \\ l \notin I_p}}^{j -1} \int_{-\infty}^{\tau} d\tau_l [H_I^{\lambda}(\tau_l),\hat{u}_I ( k_l,\tau) ] \right) \left(\prod_{\substack{l = 1 \\ l \in I_p}}^{j -1} \hat{u}_I( k_l,\tau) \right) \nn \\
\times \left( \int_{-\infty}^{\tau} d\tau_j  \int_{-\infty}^{\tau_j} d\tau' \int_{-\infty}^{\tau'} d\tau_1' \ldots \int_{-\infty}^{\tau_{p-1}'} d\tau_p' [H_I^{\lambda}(\tau_p'), \ldots [H_I^{\lambda}(\tau_1'),[H_I^{\Lambda}(\tau') ,[H_I^{\lambda}(\tau_j),\hat{u}_I( k_l,\tau)]]] \ldots] \right) \nn \\
\times \left(\prod_{\substack{l = j+1 \\ l \in I_p}}^{n} \hat{u}_I( k_l,\tau) \right) \left( \prod_{\substack{l = j+1 \\ l \notin I_p}}^{n} \int_{-\infty}^{\tau} d\tau_l [H_I^{\lambda}(\tau_l),\hat{u}_I(k_l,\tau)] \right) \ket{0}_{\rm c}. \label{n-point-comp-1}
\eea
\end{widetext}
The subscript ${\rm c}$ reminds us that we are keeping fully connected contributions only. As discussed after Eq.~(\ref{structure-u}), to obtain these contributions, we must keep only contractions including at least one operator $a_{+}$ (or $a_{+}^{\dag}$) appearing in $H^{\Lambda}_I$. To track these contractions, we have introduced the sum $\sum_{I_p}$ over the sets $I_p = 1_p, 2_p , \cdots$, consisting of all possible arrangements (of dimension $p$) of momenta labels. For example, we could write: $1_2 = (1,2)$, $2_2 = (1,3)$, $3_2 = (2,3)$, etc.

To perform the contractions appearing in (\ref{n-point-comp-1}), it is useful to define the following two quantities:
\bea
\Delta (\tau_a,\tau_b,k) = u_k(\tau_a) u_k^*(\tau_b) = v_k(\tau_a) v_k^*(\tau_b) , \\
D (\tau_a,\tau_b,k) = \Delta (\tau_a,\tau_b,k) - \Delta (\tau_b,\tau_a,k) .
\eea
In terms of these quantities, one can write the following field commutators:
\bea
\left[ \hat{u}_I( \k', \tau'),  \hat{u}_I(\k, \tau) \right] &=& (2 \pi)^3 \delta(\k' + \k) D(\tau', \tau, k) , \quad \\
\left[ \hat{v}_I( \k', \tau'),  \hat{v}_I(\k, \tau) \right] &=& (2 \pi)^3 \delta(\k' + \k) D(\tau', \tau, k)  . \quad
\eea
These relations allow us to further infer the form of the commutators involving $H_I^{\lambda}$ appearing in (\ref{n-point-comp-1}). These are found to be given by
\bea
\left[ H_I^{\lambda}(\tau'),\hat{u}_I(\k,\tau) \right] = -\dfrac{\lambda}{(\tau')^2} \dfrac{\partial [\tau' D(\tau', \tau, k)]}{\partial \tau'} \hat{v}_I(\k,\tau') ,  \qquad \\
\left[ H_I^{\lambda}(\tau'),\hat{v}_I(\k,\tau) \right] = -\dfrac{\lambda}{(\tau')^2}   D(\tau', \tau, k)  \dfrac{\partial [ \tau' \hat{u}_I(\k,\tau') ] }{\partial \tau'} . \qquad
\eea
Then, using these back into (\ref{n-point-comp-1}), and performing the relevant contractions, we find
\begin{widetext}
\bea
\langle u( \k_1,\tau) \cdots u( \k_n,\tau) \rangle_{\rm c} = i^{n+1} (2\pi)^3 \delta^{(3)}\left(\sum_{i = 1}^{n} \k_{i}\right) \dfrac{-\Lambda^4 (-1)^{n/2}}{H^4} \left(\dfrac{-\lambda H}{f} \right)^n \sum_{j = 1}^n \sum_{p = 0}^{n-1} \sum_{I_p} \sum_{{\rm perm}~q_l} \int_{-\infty}^{\tau} \!\!\!\!\!\! d\tau' \tau'^{n-4} \nn \\
  \nn \\
  \times  \int_{\tau'}^{\tau} \!\! d\tau_j \int_{-\infty}^{\tau'} \!\!\!\!\!\! d\tau_1'  \ldots \int_{-\infty}^{\tau_{p-1}'} \!\!\!\!\!\! d\tau_p' \left(\prod_{\substack{l = 1 \\ l \notin I_p}}^{j -1} \int_{-\infty}^{\tau} d\tau_l \dfrac{ \Delta(\tau_l, \tau', k_l)}{\tau_l^2} \dfrac{\partial [\tau_l D(\tau_l, \tau, k_l)]}{\partial \tau_l} \right)
 \left(\prod_{\substack{l = 1 \\ l \in I_p}}^{j -1} D(\tau_{q_l}',\tau',k_l) \dfrac{1}{\tau_{q_l}'^2} \dfrac{\partial[\tau_{q_l}' \Delta(\tau, \tau_{q_l}',k_l)]}{\partial \tau_{q_l}'} \right)
\nn \\
\times \left( \dfrac{ D(\tau', \tau_j, k_j)}{\tau_j^2} \dfrac{\partial [\tau_j D(\tau_j, \tau, k_j)]}{\partial \tau_j} \right)   \sum_{m=n/2}^{\infty} \frac{1}{(m-n/2) !} \left[ - \frac{1}{2}\left( \frac{H \tau'}{f} \right)^{2} \! \int_k \Delta( \tau' , \tau ' , k) \right]^{m-n/2}  \nn \\
\times \left(\prod_{\substack{l = j+1 \\ l \in I_p}}^{n} D(\tau_{q_l}',\tau',k_l) \dfrac{1}{\tau_{q_l}'^2} \dfrac{\partial[\tau_{q_l}' \Delta( \tau_{q_l}', \tau,k_l)]}{\partial \tau_{q_l}'} \right) \left( \prod_{\substack{l = j+1 \\ l \notin I_p}}^n \int_{-\infty}^{\tau} d\tau_l \dfrac{\Delta(\tau', \tau_l, k_l)}{\tau_l^2} \dfrac{\partial [\tau_l D(\tau_l, \tau, k_l)]}{\partial \tau_l} \right) , \quad  \label{n-point-comp-2}
\eea
\end{widetext}
if $n$ is even, and zero otherwise because the expectation value of an odd number of fields in the interaction picture vanishes identically. This is a consequence of the potential's being an even function of the isocurvature field.

In Eq.~\eqref{n-point-comp-2}, the sum over ``perm $q_l$'' refers to the sum over all possible permutations $l \to q_l$ of labels belonging to the set $I_p$. These permutations affect the arguments $\tau_{q_l}'$ appearing in some functions in the second and fourth lines of this equation.

The previous result can be simplified with the help of the following two steps: first, the sum appearing in the third line comes from the infinite loop contributions shown in Fig.~(\ref{fig:FIG_diagram_res}). They are the consequence of contractions between pairs of creation and annihilation operators $a_+^\dag$ and $a_+$ appearing in $H_I^{\Lambda}$. These terms can be resummed back into the following exponential:
\be
\sum_{m'} \frac{1}{m' !} \left[ - \frac{1}{2}\left( \frac{H \tau'}{f} \right)^{2} \! \int_k \Delta( \tau' , \tau ' , k) \right]^{m'} = e^{ - \frac{\sigma_0^2}{2 f^2} } ,
\ee
where $\sigma_0$ is nothing but the variance of $\psi$ already defined in Eq.~(\ref{sigma_0-def}). Notice that $\sigma_0$ is independent of time, and so we may factorize $\exp( - \sigma_0^2 / 2 f^2 )$ out of the $\tau'$ integral.

Secondly, in Eq.~(\ref{n-point-comp-2}) we may relabel every integration variable of the form $\tau'_i$ by a new variable $\tau_l$ (with $l \in I_p$) in such a way that, in the  functions' arguments, the $k_l$'s are always accompanied by a $\tau_l$. This relabeling allows us to recognize that the sum over all possible permutations $q_l$ becomes a sum over all domains of integration for the variables $\{\tau_l\}_{l \in I_p}$. As a result, the nested integrals of Eq.~(\ref{n-point-comp-2}) unravel, and we are led to
\begin{widetext}
\bea
\langle u(\k_1,\tau)\cdot \cdot \cdot u(\k_n,\tau) \rangle_{\rm c} = i^{n+1} (2\pi)^3 \delta^{(3)}\left(\sum_{i = 1}^{n} \k_{i}\right) \dfrac{-\Lambda^4 (-1)^{n/2}}{H^4} \left(\dfrac{-\lambda H}{f} \right)^n e^{-\frac{\sigma_0^2}{2f^2}} \sum_{j = 1}^n \sum_{p = 0}^{n-1} \sum_{I_p} \int_{-\infty}^{\tau} d\tau' \tau'^{n-4} \nn \\
\times \left(\prod_{\substack{l = 1 \\ l \in I_p}}^{j -1} \int_{-\infty}^{\tau'} d\tau_l  D(\tau_{l},\tau',k_l) \dfrac{1}{\tau_{l}^2} \dfrac{\partial[\tau_{l} \Delta(\tau, \tau_{l},k_l)]}{\partial \tau_{l}} \right) \left(\prod_{\substack{l = 1 \\ l \notin I_p}}^{j -1} \int_{-\infty}^{\tau} d\tau_l \dfrac{ \Delta(\tau_l, \tau', k_l)}{\tau_l^2} \dfrac{\partial [\tau_l D(\tau_l, \tau, k_l)]}{\partial \tau_l} \right) \nn \\
\times  \left(\int_{\tau'}^{\tau} d\tau_j \dfrac{ D(\tau', \tau_j, k_j)}{\tau_j^2} \dfrac{\partial [\tau_j D(\tau_j, \tau, k_j)]}{\partial \tau_j} \right) \nn \\
\times \left(\prod_{\substack{l = j+1 \\ l \in I_p}}^{n} \int_{-\infty}^{\tau'} d\tau_l  D(\tau_{l},\tau',k_l) \dfrac{1}{\tau_{l}^2} \dfrac{\partial[\tau_{l} \Delta( \tau_{l}, \tau,k_l)]}{\partial \tau_{l}} \right) \left( \prod_{\substack{l = j+1 \\ l \notin I_p}}^n \int_{-\infty}^{\tau} d\tau_l \dfrac{\Delta(\tau', \tau_l, k_l)}{\tau_l^2} \dfrac{\partial [\tau_l D(\tau_l, \tau, k_l)]}{\partial \tau_l} \right). \label{n-point-comp-3}
\eea
The previous expression can be simplified further by performing the summation over the index $p$ (including the sum over the sets $I_p$ introduced earlier). Notice that these sums allow one to rewrite the second and fourth lines of Eq.~(\ref{n-point-comp-3}) as the multiplication of pairs of terms
\bea
\langle u(\k_1,\tau)\cdot \cdot \cdot u(\k_n,&&\tau) \rangle_{\rm c} = i^{n+1} (2\pi)^3 \delta^{(3)}\left(\sum_{i = 1}^{n} \k_{i}\right) \dfrac{-\Lambda^4 (-1)^{n/2}}{H^4} \left(\dfrac{-\lambda H}{f} \right)^n e^{-\frac{\sigma_0^2}{2f^2}} \sum_{j = 1}^n  \int_{-\infty}^{\tau} d\tau' \tau'^{n-4} \nn \\
&& \times  \prod_{\substack{l = 1}}^{j-1} \left(\int_{-\infty}^{\tau'} \dfrac{d\tau_l}{\tau_{l}^2}  D(\tau_{l},\tau',k_l) \dfrac{\partial[\tau_{l} \Delta(\tau, \tau_{l},k_l)]}{\partial \tau_{l}} + \int_{-\infty}^{\tau} \dfrac{d\tau_l}{\tau_{l}^2}  \Delta(\tau_l, \tau', k_l) \dfrac{\partial [\tau_l D(\tau_l, \tau, k_l)]}{\partial \tau_l} \right) \nn \\
&& \times  \left(\int_{\tau'}^{\tau} \dfrac{d\tau_j}{\tau_{j}^2} D(\tau', \tau_j, k_j) \dfrac{\partial [\tau_j D(\tau_j, \tau, k_j)]}{\partial \tau_j} \right) \nn \\
&& \times  \prod_{\substack{l = j+1}}^{n} \left(\int_{-\infty}^{\tau'} \dfrac{d\tau_l}{\tau_{l}^2}  D(\tau_{l},\tau',k_l) \dfrac{\partial[\tau_{l} \Delta( \tau_{l},\tau,k_l)]}{\partial \tau_{l}} + \int_{-\infty}^{\tau} \dfrac{d\tau_l}{\tau_{l}^2} \Delta(\tau',\tau_l, k_l) \dfrac{\partial [\tau_l D(\tau_l, \tau, k_l)]}{\partial \tau_l} \right) \!. \label{n-point-comp-4}
\eea
\end{widetext}
We can now rewrite this expression in a more compact way by noticing that the propagators of the $\psi$ and $\zeta$ fields are given by (up to a \(\sqrt{2 \epsilon}\) factor for $\zeta$)
\be
\bar{\Delta}(\tau',\tau_l, k_l) = H^2 \tau'\tau_{l}\Delta(\tau',\tau_l, k_l).
\ee
By switching to $\zeta,\psi$ variables the $\tau'^n$ factor from the first line of \eqref{n-point-comp-4} and the $\tau^n$ factor, coming from the conversion of the external fields, combine with the $\Delta$ propagators to form the $\bar{\Delta}$ ones. Then we may conveniently introduce the following set of functions
\bea
\mathcal{G}_{+}(k_l,\tau',\tau) \equiv \frac{(-i\lambda)}{H^3} \int_{-\infty}^{\tau} \dfrac{d\tau_l}{\tau_{l}^3}  \left(\bar{\Delta}(\tau_l, \tau', k_l) \dfrac{\partial [\bar{D}(\tau_l, \tau, k_l)]}{\partial \tau_l}   \right. \nn \\
\left. +  \bar{D}(\tau_{l},\tau',k_l) \dfrac{\partial[\bar{\Delta}(\tau, \tau_{l},k_l)]}{\partial \tau_{l}} \Theta(\tau'-\tau_l)  \right), \,\,\,\,\,\,\,\,\,\,\,\,\,\,\,\,\,\,\, \\
\mathcal{G}_{-}(k_l,\tau',\tau) \equiv \frac{(-i\lambda)}{H^3} \int_{-\infty}^{\tau} \dfrac{d\tau_l}{\tau_{l}^3}  \left( \bar{\Delta}(\tau',\tau_l, k_l) \dfrac{\partial [\bar{D}(\tau_l, \tau, k_l)]}{\partial \tau_l} \right. \nn \\
\left. + \bar{D}(\tau_{l},\tau',k_l) \dfrac{\partial[\bar{\Delta}( \tau_{l},\tau,k_l)]}{\partial \tau_{l}} \Theta(\tau'-\tau_l)  \right), \,\,\,\,\,\,\,\,\,\,\,\,\,\,\,\,\,\,\,
\eea
where $\Theta(\tau)$ is the usual Heaviside function. The functions $\mathcal{G}_{\pm}(k_l,\tau',\tau)$ are nothing but the \emph{mixed propagators} defined in Ref.~\cite{Chen:2017ryl}, with the external -- or boundary, in the language of \cite{Chen:2017ryl} -- time, denoted by $\tau$ here, left explicit as an argument. Moreover, subtracting them we find
\bea
\mathcal{G}_{+}(k_j,\tau',\tau) - \mathcal{G}_{-}(k_j,\tau',\tau) =  \nn \\
\frac{(+i\lambda)}{H^3} \int_{\tau'}^{\tau} \dfrac{d\tau_j}{\tau_{j}^3} \bar{D}(\tau', \tau_j, k_j) \dfrac{\partial [ \bar{D}(\tau_j, \tau, k_j)]}{\partial \tau_j}.
\eea
Replacing these definitions back into Eq.~(\ref{n-point-comp-4}) and noting that consecutive terms with alternating signs cancel out in the sum over $j$, we end up with the following simple expression for the $n$-point correlation function
\bea
\langle \zeta(\k_1,\tau)\cdot \cdot \cdot \zeta(\k_n,\tau) \rangle_{\rm c} = \dfrac{i\Lambda^4}{H^4} (2\pi)^3 \delta^{(3)}\left(\sum_{i = 1}^{n} \k_{i}\right)  e^{-\frac{\sigma_0^2}{2f^2}} \nn \\
 \left(-\dfrac{H^2}{2 \epsilon f^2} \right)^{n/2} \int_{-\infty}^{\tau} \frac{d\tau'}{\tau'^{4}} \Big[ \mathcal{G}_{+}(k_1,\tau',\tau) ... \mathcal{G}_{+}(k_n,\tau',\tau)  \nn \\
  - \mathcal{G}_{-}(k_1,\tau',\tau) ... \mathcal{G}_{-}(k_n,\tau',\tau) \Big]. \,\,\,\,\,\,\,\,\,\,\,\,\,\,\,\,\,\,\, \label{n-point-comp-G}
\eea
It is rewarding to notice that this result is exactly what we would have obtained had we used the diagrammatic rules of \cite{Chen:2017ryl}, after adequately treating the loop contributions.

Let us now perform the integrals of Eq.~(\ref{n-point-comp-G}) [or, equivalently, Eq.~(\ref{n-point-comp-4})] to obtain a simple and useful expression for the $n$-point correlation function in momentum space.  As discussed in Sec.~\ref{sec:linearth}, recall that the effect of the $H_I^{\lambda}$ is to source the evolution of the amplitude of $\zeta$ (or, equivalently, $u$) on superhorizon scales. Thus, for a given fixed set of $k_l$'s, the integration domain of each integral appearing in Eq.~(\ref{n-point-comp-G}) can be split into two parts. Before horizon crossing, where $|\tau_l| k_l > 1$ (and $|\tau'| k_{l} > 1$) and after horizon crossing, where the opposite is true: $|\tau_l| k_l < 1$ (and $|\tau'| k_{l} < 1$). The integrants are highly oscillatory in the domain $|\tau_l| k_l > 1$ (and $|\tau'| k_{l} > 1$). These oscillatory contributions would have vanished had we kept track of the $\epsilon$ prescription. And so, we may simply disregard the contributions to (\ref{n-point-comp-4}) coming from these domains.

On the other hand, the integration of these functions over the domain $|\tau_l| k_l < 1$ (and $|\tau'| k_{l} < 1$)  gives us expressions that dominate if the upper limit $\tau$ is such that $|\tau| k_l \ll 1$ (in fact, the integration over the domain  $|\tau_l| k_l < 1$ diverges as $\tau \to 0$). Thus, instead of obtaining exact expressions for these integrals, we may seek the infrared divergent contributions that dominate on superhorizon scales. To do this explicitly for each integral, we introduce an arbitrary time $\tau_0$ such that $|\tau_0| k_l < 1$ for all $k_l$'s. We will use $\tau_0$ to cut off the lower limit of every time integral. The upper limit may be either the end of inflation or some other value, depending on the exact mass of the scalar field. In either case, we assume that the amount of e-folds that different modes spend outside the horizon is approximately the same, as discussed in the main text.

Then, within the integration domains $\tau_l' \in (\tau_0, \tau')$, one may simplify a few of the integrated functions as
\bea
 \dfrac{D(\tau_{l},\tau',k_l)}{\tau_{l}^2} \dfrac{\partial[\tau_{l} \Delta( \tau_{l}, \tau,k_l)]}{\partial \tau_{l}}
\simeq  \dfrac{i e^{-i k_l \tau_l}}{6 k_l \tau_l \tau} \!\! \left[\dfrac{\tau'^2}{\tau_l} - \dfrac{\tau_l^2}{\tau'} \right] , \,\,\, \qquad  \\
 \dfrac{D(\tau_{l},\tau',k_l)}{\tau_{l}^2} \dfrac{\partial[\tau_{l} \Delta( \tau, \tau_{l},k_l)]}{\partial \tau_{l}} \simeq \dfrac{i e^{+i k_l \tau_l}}{6 k_l \tau_l \tau}  \!\! \left[\dfrac{\tau'^2}{\tau_l} - \dfrac{\tau_l^2}{\tau'} \right] . \,\,\, \qquad
\eea
Then, it is direct to find
\bea
\int_{\tau_0}^{\tau'} \!\!\!\!\! d\tau_l \, D(\tau_{l},\tau',k_l) \dfrac{1}{\tau_{l}^2} \dfrac{\partial[\tau_{l} \Delta( \tau_{l}, \tau,k_l)]}{\partial \tau_{l}} \qquad \nn \\
 \simeq \dfrac{i e^{-i k_l \tau'}}{6 k_l \tau} \left(\dfrac{\tau'^2}{\tau_0} - \dfrac{3 \tau'}{2} + \dfrac{\tau_0^2}{2 \tau'} \right).
\eea
Similarly, the other relevant integrals may be evaluated as
\bea
\int_{\tau_0}^{\tau'} \!\!\!\!\! d\tau_l \, D(\tau_{l},\tau',k_l) \dfrac{1}{\tau_{l}^2} \dfrac{\partial[\tau_{l} \Delta(  \tau,\tau_{l},k_l)]}{\partial \tau_{l}} \nn \,\,\,\,\,\,\,\,\,\, \\
 \simeq \dfrac{i e^{+i k_l \tau'}}{6 k_l \tau} \left(\dfrac{\tau'^2}{\tau_0} - \dfrac{3 \tau'}{2} + \dfrac{\tau_0^2}{2 \tau'} \right), \qquad  \\
\int_{\tau_0}^{\tau} \!\!\!\! d\tau_l \,  \dfrac{u_{k_l}(\tau_l)}{\tau_{l}^2} \dfrac{\partial[\tau_{l} D( \tau_{l}, \tau,k_l)]}{\partial \tau_{l}}  \simeq \dfrac{2\ln(\tau_0/\tau)}{(2 k_l)^{3/2} \tau} , \,\,\, \qquad  \\
\int_{\tau_0}^{\tau} \!\!\!\! d\tau_l \,  \dfrac{u_{k_l}^*(\tau_l)}{\tau_{l}^2} \dfrac{\partial[\tau_{l} D( \tau_{l}, \tau,k_l)]}{\partial \tau_{l}}  \simeq - \dfrac{2\ln(\tau_0/\tau)}{(2 k_l)^{3/2} \tau} , \,\,\, \qquad \\
\int_{\tau'}^{\tau} \!\!\!\! d\tau_j \,  \dfrac{D(\tau',\tau_j,k_j)}{\tau_{j}^2} \dfrac{\partial[\tau_{j} D( \tau_{j}, \tau,k_l)]}{\partial \tau_{j}} \nn \,\,\,\,\,\,\,\,\,\, \\
 \simeq  \dfrac{2\ln(\tau'/\tau)}{(2 k_j)^{3/2} \tau} \sqrt{\dfrac{2}{k_j}} \dfrac{k_j^2 \tau'^2}{3} .\,\,\,  \qquad
\eea
Using these results back into (\ref{n-point-comp-4}), and performing the final integral over \(\tau'\), we finally obtain
\bea
&& \langle u(\k_1,\tau)\cdot \cdot \cdot u(\k_n,\tau) \rangle_{\rm c} \simeq (-1)^{n/2} (2\pi)^3 \delta^{(3)}\left(\sum_{i = 1}^{n} \k_{i}\right)  \quad \nn \\
&& \,\, \times \dfrac{\Lambda^4}{H^4} e^{-\frac{\sigma_0^2}{2f^2}} \left(\dfrac{\lambda H}{2 f \tau} \right)^n \dfrac{1}{3}  \dfrac{k_1^3 + ... + k_n^3}{k_1^3 \cdot \cdot \cdot k_n^3} \text{ln}\left(\dfrac{\tau_0}{\tau}\right)^{n+1}. \quad \label{n-point-comp-5}
\eea
Let us remind the reader that the previous equation holds as written for even $n$ only. If $n$ is odd, the correlator vanishes identically because the potential is an even function of the isocurvature field. This result differs by a factor $1/2$ from the Gaussian relation shown in Eq.~(\ref{zeta-n-psi-n}) based on the linear evolution of $\zeta$ at superhorizon scales. This factor appears as a consequence of the integration
\be
\int_{\tau_0}^{\tau} \!\! d\tau' \, \dfrac{\text{ln}(\tau'/\tau)}{\tau'} = - \dfrac{1}{2}\text{ln}^2 (\tau_0/\tau) ,
\ee
and may be understood as a nonlinear effect due to two classes of vertices involved in the computation. The final expression offered in (\ref{n-point-comp-5}) is the main result of this section, and leads directly to Eq.~(\ref{n-point-final-res-zeta}).

Given that $\tau_0$ is chosen in such a way that $|\tau_0| k_l < 1$ (for all $k_l$'s), then we may roughly identify the quantity $\ln (\tau_0 /\tau)$ as the number of $e$-folds after all the modes left the horizon:
\be
\Delta N \simeq \ln (\tau_0/\tau) .
\ee
Then, it is possible to verify that the infrared contributions to the integrals leading to (\ref{n-point-comp-5}) dominate when the condition
\be
 \Delta N \gg  1
\ee
is satisfied. Recall that, in order to deal with the system perturbatively, our computation assumed $\lambda \ll 1$. This means that only after a time $\Delta N \sim 1/\lambda$ the effects (both linear and nonlinear) induced by $\psi$ start to take over, as evidenced by the computations of Sec.~\ref{sec:linearth}.

\section{Structure of $I_n$} \label{app_about_I_n}

Here we derive a few properties of $I_n$ defined in Eq.~(\ref{full-I-n}). To start with, notice that the sum $k_1^3 + \cdots + k_n^3$ appearing in Eq.~(\ref{full-I-n}) leads to $n$ identical integrals. Then, by writing the Dirac delta function as $\delta (\p) = (2 \pi)^{-3} \int_r e^{- i \p \cdot \r}$, the integral $I_n$ becomes
\be
I_n = \frac{n}{(2 \pi)^3}   \int_r  \left[ \int_{k < k_L} \!\!\!\!\! \frac{e^{- i \k \cdot \r} }{k^3 } \right]^{n-1} \int_{q < k_L}   \!\!\!\!\!  e^{- i \q \cdot \r} .
\ee
Let us recall that we are also dealing with an IR cutoff $k_{\rm IR}$. The previous expression may be simplified by first redefining the integration variables as
\be
\r = \x / k_L, \qquad   \k = \y k_L , \qquad \q = \z k_L  ,
\ee
and then by solving the angular parts of the $\x$, $\y$ and $\z$ integrals. These two steps lead to
\bea
I_n (\xi) &=& \frac{n}{(2 \pi^2)^{n+1}}  \int_0^{\infty} \frac{dx}{x} \left[ \int^{1}_{\xi^{-1}} \frac{d y}{y} \frac{\sin (y x)}{y x} \right]^{n-1}  \nn \\
&& \times  \int^{1}_{\xi^{-1}} \!\!\! d z \, z \,  x^2 \sin (z x)  , \label{I-n-x-beta-0}
\eea
where $\xi =  k_L/ k_{\rm IR}$ is the ratio of scales available to observers introduced in Sec.~\ref{subsec:n-point}. In Eq.~(\ref{I-n-x-beta-0}), we have emphasized that $I_n$ is a function of $\xi$. It may be noticed that this function satisfies
\be
I_n(1) = 0 .
\ee
Next, the $y$ and $z$ integrals may be solved to give
\be
I_n (\xi) = \frac{n }{(2 \pi^2)^{n+1}} \int_0^{\infty} \!\! \frac{dx}{x} G(\xi,x) \left[ F(\xi , x) \right]^{n-1}   ,  \quad \label{app-full-I_n} 
\ee
where we have introduced the functions
\bea
G(\xi , x) &=&  \sin (x) - x \cos (x) - \sin (x / \xi)  \nn \\ 
&&  + (x / \xi) \cos (x / \xi)  , \qquad \\
F(\xi , x) &=& {\rm Ci} (x) - \frac{\sin (x)}{x}  - {\rm Ci} ( x / \xi ) + \frac{\sin ( x / \xi )}{ x / \xi} . \qquad
\eea
Here, $ {\rm Ci} (x)$ is the cosine integral function.  It is hard to obtain a simple analytical representation of $I_n$ (for arbitrary $n$) beyond that shown in Eq.~(\ref{app-full-I_n}). In the particular case $n=2$, one finds (for $ \xi \geq 1$)
\be
I_2 (\xi)  = \frac{\pi}{(2 \pi^2)^3} \ln \xi .
\ee
In the cases $n=3$ and $n=4$, we are able to obtain exact analytical expressions but it is more useful to write down the results in the limit $\xi \gg 1$:
\bea
I_3 (\xi)  &=& \frac{3 \pi}{2 (2 \pi^2)^{4}}  \left[  \left( \ln \xi \right)^2 + 1 -  \frac{\pi^2}{6}  \right] + \mathcal O(\xi^{-1}), \qquad  \\
I_4 (\xi)  &=& \frac{4 \pi}{2 (2 \pi^2)^{5}} \Bigg[  \left( \ln \xi \right)^3 - \frac{\pi^2 - 3}{4}   \ln \xi  \nn \\
&& + \frac{11}{4} \zeta (3) - \frac{43}{24} \Bigg] + \mathcal O(\xi^{-1}) . \label{I_4-direct}
\eea
While for $n \geq 5$ we are not able to obtain simple general expressions in the large $\xi$ limit, we can still derive a useful property about $I_n$. Indeed, it is possible to show that $I_n$ has the following asymptotic form for $n \geq 4$:
\be
I_n =  I_n^0 + \Delta I_n , \label{I-n-expansion-0}
\ee
where 
\bea
I_n^0 &=& \frac{n \pi}{2 (2 \pi^2)^{n+1}}   \left( \ln \xi \right)^{n-1} ,  \label{I-n-expansion-1} \\
\Delta I_n &=& - I_n^0 \times C \frac{(n-1) (n-2)}{2} \left( \ln \xi \right)^{-2} \nn \\
&& + \mathcal O \left[ (\ln \xi )^{-3} \right] ,  \label{I-n-expansion-2}
\eea
with $C =(\pi^2-3)/12 \simeq 0.6$. 

Before deriving Eqs.~(\ref{I-n-expansion-0})-(\ref{I-n-expansion-2})  let us briefly point out their importance: even for very large values of $\xi$, the difference between the leading term $I_n^0$ and the correction $\Delta I_n$ is given by a factor $\left( \ln \xi \right)^2$, which is not necessarily large enough. In fact, the result shown in Eq.~(\ref{I-n-expansion-2}) implies that $\Delta I_n$ will quickly become of order $I_n^0$ as $n$ grows. It is a simple matter to verify that if $\ln \xi = 60$, then $\Delta I_n$ will reach one-tenth of $I_n^0$ around $n\simeq 35$. Nevertheless, the important question here is to what extent the correction $\Delta I_n$ modifies the PDF derived in Sec.~\ref{subsec:PDF-derivation-from-n-points}, obtained under the assumption that $I_n$ can be taken as $I_n^0$. The answer turns out to be rather simple: by performing the same reconstruction carried out in Sec.~\ref{subsec:PDF-derivation-from-n-points}, using the ansatz~(\ref{PDF-ansatz}), we find that $\Delta I_n$ implies an extra contribution to~(\ref{PDF-1}) that has an oscillatory behavior set by the scale $f_{\zeta}$, as expected. This time, however, the oscillations come with different factors in their amplitudes. Among these terms we find a factor of $\sigma_L^4/f^4$ (or, equivalently, $\sigma_{\zeta}^4/f_{\zeta}^4$), and thus comparing with the terms in the uncorrected PDF, the leading of which is $\sigma_L^2/f^2$, we get an estimate of how small $f$ can be so that Eq.~(\ref{PDF-1}) remains accurate. The final answer is that in order to be able to neglect the correction to the PDF, at the very least we must have
\be
f \gtrsim \frac{\sigma_L}{\sqrt{\ln \xi}}    .  \label{f-log-cutoffs}
\ee
This result severely limits the applicability of the PDF of Eq.~(\ref{PDF-1}), and forces us to consider the more general reconstruction carried out in Sec.~\ref{subsec:PDF-derivation-from-n-points-full}.

\subsection{Derivation of $\Delta I_n$}

To show Eqs.~(\ref{I-n-expansion-0})-(\ref{I-n-expansion-2}), we may split the integral (\ref{I-n-x-beta-0}) into two parts, one containing the term $\sin(x) - x \cos (x)$ and the other containing the term $- \sin(x / \xi) + (x / \xi) \cos ( x / \xi)$. Then by changing the integration variable of the second part as $x \to x / \xi$, we end up with the following expression
\be
I_n (\xi) = \frac{ n \pi}{2 (2 \pi^2)^{n+1}}  \left[ \bar I_n (\xi)  + (-1)^n  \bar I_n (1 / \xi)  \right], \label{I-n-x-beta-v2}
\ee
where we have defined $ \bar I_n (\xi)$ as
\be
\bar I_n (\xi) \equiv  \frac{2}{\pi} \int_0^{\infty} \frac{dx}{x} \left[ \sin(x) - x \cos (x)  \right]   \left[ F (\xi, x) \right]^{n-1}   .  \qquad
\ee
Notice that for large $\xi$ the second contribution to (\ref{I-n-x-beta-v2}), given by $\bar I_n(1/\xi)$, is very suppressed compared to the first one, given by $\bar I_n(\xi)$. To see this, it is enough to see that $F(1/\xi , x)$ is a function that becomes quickly independent of $\xi$ for values $x > 1/ \xi$, and so one finds that in the relevant integration domain the function is essentially given by
\be
F(1 / \xi, x) \simeq  {\rm Ci} (x) - \frac{\sin (x)}{x} .
\ee
This implies that the contribution to (\ref{I-n-x-beta-v3}) coming from $\bar I_n(1/\xi)$ is at most of order 1. Then, we are left with
\be
I_n (\xi) = \frac{n \pi}{2 (2 \pi^2)^{n+1}}  \bar I_n (\xi) + \mathcal O (1)  . \label{I-n-x-beta-v3}
\ee
Next, notice that the function $F(\xi,x)$ that appears inside the integral $\bar I_n (\xi) $ satisfies
\be
F(\xi, 0) = \ln \frac{1}{\xi} .
\ee
In addition, in the range $1 < x  \leq \xi$, it is well approximated by $F_\xi(x)  \simeq 1 - \gamma +  \ln \frac{\xi}{ x } $, whereas, for values $x > \xi$, the function $F_\xi(x) $ vanishes quickly. We can therefore write
\be
F(\xi,x)  \simeq 1 - \gamma +  \ln \frac{\xi}{ x }  + \epsilon (\xi x), \label{approx-F}
\ee
where $\gamma$ is the Euler-Mascheroni constant and $\epsilon (y)$ is a slowly varying function that remains small in the relevant interval $1 < x  \leq \xi$. In fact, this function is suppressed everywhere $1 < x  \leq \xi$ and its largest value is of order $0.1$ when $x / \xi  = 1$. It is enough to take $\epsilon (y) = y^2 / 12$. Then, we can write
\bea
\bar I_n (\xi)  =  \frac{2}{\pi} \int_0^{x_*} \frac{dx}{x} \left[ \sin(x) - x \cos (x)  \right]  \nn \\
\left[ 1 - \gamma +  \ln \frac{\xi }{x }  + \epsilon (\xi x) \right]^{n-1}  ,
\eea
where $x^* \gtrsim \xi$ has been introduced to cut off the integral, since for values larger than $x^*$ the function $F(\xi , x)$ is highly suppressed. The introduction of $x_*$ has the benefit of allowing us to use approximation (\ref{approx-F}) inside the integral.  Now, taking a derivative with respect to $\xi$, we find
\bea
\bar I_n ' (\xi)  =   (n-1)  \frac{2}{\pi} \int_0^{x_*} \frac{dx}{x} \left[ \sin(x) - x \cos (x)  \right]  \nn \\ \times
\left[  \frac{1}{\xi} + \epsilon ' (\xi x) x \right]
\left[ 1 - \gamma +  \ln \frac{1}{\xi x }  + \epsilon (\xi x) \right]^{n-2}  . \quad
\eea
This equation leads to
\bea
\bar I_n ' (\xi)  &=&  \frac{(n - 1)}{\xi} \bar I_{n-1}  (\xi) \nn \\
&& +  (n-1)  \frac{2}{\pi} \int_0^{x_*} \frac{dx}{x} \left[ \sin(x) - x \cos (x)  \right]
  \nn \\
 &&
 \times \epsilon ' (\xi x) x \left[ 1 - \gamma +  \ln \frac{1}{\xi x }  + \epsilon (\xi x) \right]^{n-2} .
\eea
Because $\xi x \epsilon ' (\xi x)$ is of order $\epsilon$ in the entire domain $1 < x  \xi$, the second term is clearly subleading with respect to the first one. Then, we can simply disregard the second term, and write
\be
\bar I_n ' (\xi)  = \frac{(n - 1)}{\xi} \bar I_{n-1}  (\xi) .
\ee
This relation allows us to obtain $\bar I_{n} (\xi)$ from $ \bar I_{n-1}  (\xi)$. For instance, by direct computation, we are able to deduce that in the large $\xi$ limit, $\bar I_3(\xi)$ and $\bar I_3(1/\xi)$ are given by
\bea
\bar I_{3} (\xi) &=& ( \ln \xi )^2 - \frac{\pi^2 - 3}{12} + \mathcal O(\xi^{-1}) , \\
\bar I_{3} (1/\xi) &=&  \frac{\pi^2 - 9}{12} + \mathcal O(\xi^{-1}) .
\eea
From this result, we identify $C = (\pi^2 - 3) / 12$. Then, we have
\be
\bar I_4 ' (\xi)  =  \frac{3}{\xi} \left( [ \ln (1/ \xi) ]^2 - C \right) .
\ee
Solving this relation, we end up with
\be
\bar I_4  (\xi)  = ( \ln   \xi )^3 - 3 C \ln \xi + \mathcal O(1) .
\ee
Notice that this result correctly accounts for the form of $I_4$ previously shown in Eq.~(\ref{I_4-direct}). Repeating this step many times, we end up with the following general expression for $\bar I_n  (\xi)$:
\be
\bar I_n  (\xi)  =  ( \ln \xi )^{n-1}  -  C \frac{(n-1)(n-2)}{2} ( \ln \xi )^{n-3} ,
\ee
from where Eqs.~(\ref{I-n-expansion-0})-(\ref{I-n-expansion-2}) follow.

\end{appendix}

\end{document}